\pdfoutput=1
\documentclass[aps,prb,reprint,showpacs,twocolumn,superscriptaddress,preprintnumbers,amsmath,epsfig, floatfix]{revtex4-2}
\usepackage{graphicx}
\usepackage{amsmath}
\usepackage[version=4]{mhchem}
\usepackage{adjustbox}
\usepackage{chemformula}
\usepackage{placeins}
\usepackage{amssymb}
\usepackage{amsfonts}
\usepackage{dcolumn}
\usepackage{dsfont}
\usepackage{latexsym}
\usepackage{rotating}
\usepackage{color}
\usepackage{latexsym}
\usepackage{bbm}
\usepackage{siunitx}
\usepackage{subfigure}
\usepackage{float}
\usepackage{epsfig}
\usepackage{psfrag}
\usepackage{natbib}
\usepackage{bm}
\usepackage{booktabs}
\usepackage{amsthm}
\usepackage{eucal}
\usepackage{mathrsfs}   
\usepackage{array}      
\usepackage{float}
\usepackage{url}
\usepackage{times}
\usepackage{mathptmx}
\usepackage{braket}
\usepackage{mathtools}
\usepackage{amsmath,amssymb}
\usepackage{epsfig}
\usepackage{amsmath,amssymb}
\usepackage{hyperref}
\usepackage{hyperref}
\usepackage[all]{hypcap}
\usepackage{multirow}
\usepackage{color}

\hypersetup{
colorlinks=true,final=true,
        linkcolor=blue,
        citecolor=blue,
        filecolor=blue,
        urlcolor=blue,
}

\usepackage{dcolumn}
\usepackage{bm}
\usepackage{natbib}
\usepackage{times}
\usepackage{float}
\usepackage[utf8]{inputenc}
\usepackage[english]{babel}
\usepackage{keyval}
\graphicspath{{figs/}{figs:}{\figs}}
\setkeys{Gin}{width=0.90\columnwidth}
\newcommand{\comment}[1]{\textcolor{red}{#1}}
\renewcommand{\comment}[1]{\relax}

\newcommand{\todelete}[1]{\textcolor{green}{\sout{#1}}}
\renewcommand{\todelete}[1]{\relax}

\usepackage {romannum}

\begin{document}

\title{Antisite-disorder driven tuning of magnetic properties and exchange-bias in Nd$_{2-x}$Sr$_{x}$CoMnO$_{6-\delta}$ $(0 \leq x \leq 1)$ ($\delta \sim 0.5$) double perovskites}

\date{\today}
\author{Kazi Parvez Islam}
\affiliation{Department of Physics, Indian Institute of Technology Kharagpur, Kharagpur 721302, West Bengal, India}

\author{Jayjit Kumar Dey}
\affiliation{Deutsches Elektronen-Synchrotron DESY, Notkestr. 85, 22607 Hamburg, Germany}

\author{Sourav Chowdhury}
\affiliation{Deutsches Elektronen-Synchrotron DESY, Notkestr. 85, 22607 Hamburg, Germany}

\author{Samyabrata Paria}
\affiliation{Department of Physics, Indian Institute of Technology Kharagpur, Kharagpur 721302, West Bengal, India}

\author{Flora Banerjee}
\affiliation{Department of Chemistry, Indian Institute of Technology Kharagpur, Kharagpur 721302, West Bengal, India}

\author{Suryakanta Mishra}
\affiliation{Department of Physics, Indian Institute of Technology Kharagpur, Kharagpur 721302, West Bengal, India}

\author{Suman Kalyan Samanta}
\affiliation{Department of Chemistry, Indian Institute of Technology Kharagpur, Kharagpur 721302, West Bengal, India}

\author{Moritz Hoesch}
\affiliation{Deutsches Elektronen-Synchrotron DESY, Notkestr. 85, 22607 Hamburg, Germany}

\author{Robert Dankelman}
\affiliation{TU Delft Reactor Institute, Mekelweg 15, Delft, The Netherlands}

\author{Indu Dhiman}
\affiliation{TU Delft Reactor Institute, Mekelweg 15, Delft, The Netherlands}

\author{Debraj Choudhury}
\email{debraj@phy.iitkgp.ac.in}
\affiliation{Department of Physics, Indian Institute of Technology Kharagpur, Kharagpur 721302, West Bengal, India}

\begin{abstract}
\noindent 
We demonstrate precise control of exchange bias (EB) in the Nd$_{2-x}$Sr$_x$CoMnO$_{6-\delta}$ ($0 \leq x \leq 1$) double-perovskite series through Sr$^{2+}$ induced hole doping, unveiling a remarkable transition between normal and inverse EB states. Employing neutron powder diffraction and X-ray absorption spectroscopy, we reveal a structural evolution from a B-site-ordered monoclinic ($P2_1/n$) phase to a disordered rhombohedral ($R\overline{3}c$) phase with increasing $x$, accompanied by a shift in the effective Co valence from +2 toward +3, while the Mn valence remains essentially unchanged. DC magnetization measurements indicate a gradual suppression of ferromagnetism with hole doping, whereas AC susceptibility measurements at $x = 0.75$ reveal pronounced cluster-glass behavior and the highest EB field of $\sim 4$ kOe at 8 K under a 6 T cooling field. After correcting for minor-loop effects, we identify robust inverse EB at $x = 0.75$, persisting even under a cooling field of 6 T. We attribute this phenomenon to competing ferromagnetic--antiferromagnetic and ferromagnetic--glassy interfaces, governed by strong magnetic frustration and the magnetocrystalline anisotropy associated with rare-earth 4$f$ electrons. These findings elucidate the pivotal role of doping-induced structural and magnetic competition in tailoring EB behavior in rare-earth double perovskites, providing new insights for the design of advanced magnetic materials.
\end{abstract}
\maketitle
\section{Introduction}
\begin{figure}
\scalebox{0.5}
{\includegraphics[width=\textwidth]{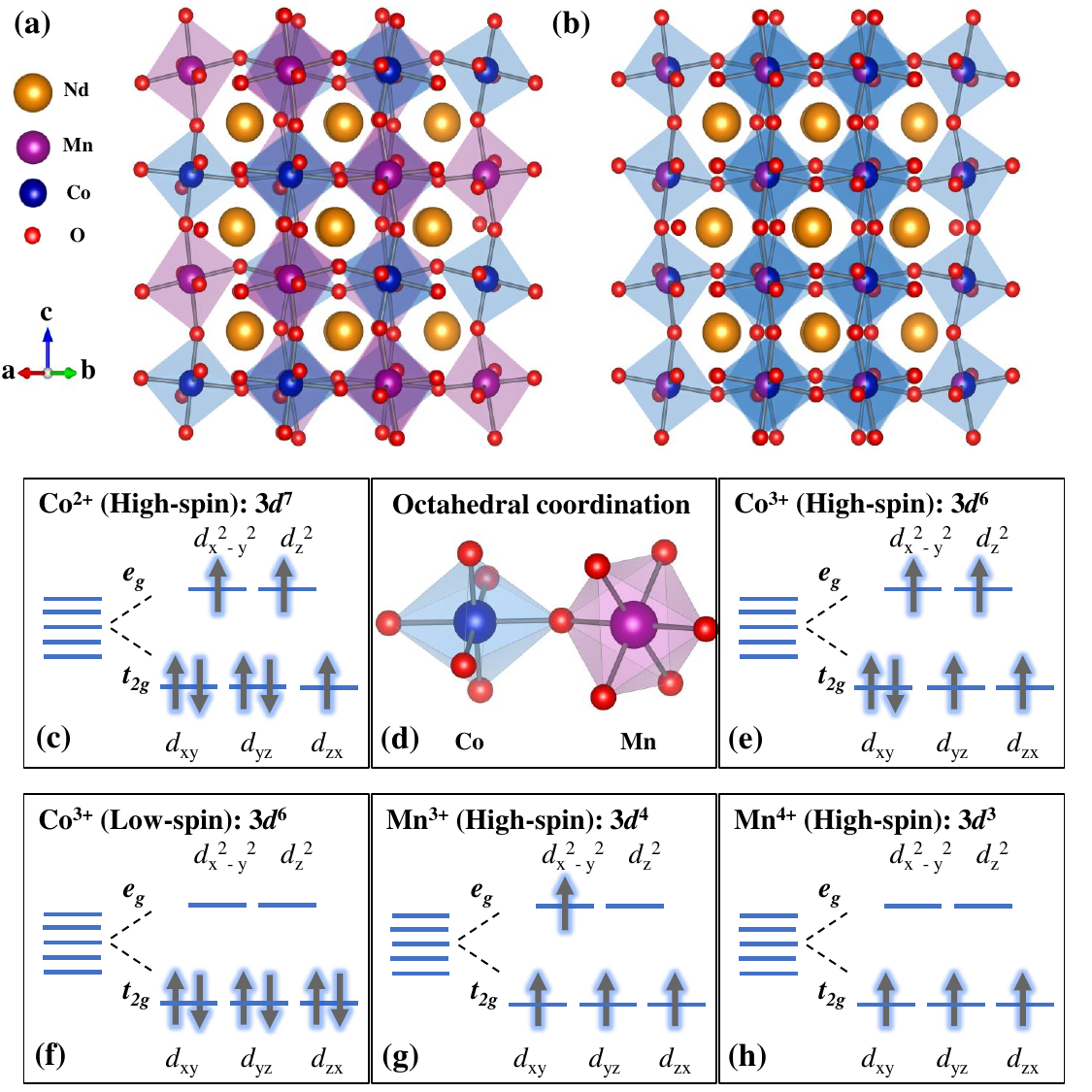}}
\caption{(color online)  (a) Crystal structure of double perovskite Nd$_{2}$CoMnO$_{6}$ with different atoms highlighted. Notice the alternate stacking of Co/Mn atoms along three crystallographic directions which leads to the B-site ``ordered" structure. (b) B-site ``disordered" crystal structure where each B-site is occupied by either Co or Mn with equal probability. (c), (e), (f), (g) and (h) highlight different electronic configurations of Co$^{2+}$ (high-spin), Co$^{3+}$ (high-spin), Co$^{3+}$ (low-spin), Mn$^{3+}$ (high-spin) and Mn$^{4+}$ (high-spin) respectively. (d) shows octahedral coordination of Co and Mn in these compounds.}\label{Fig. (1)}
\end{figure}
Interest in double perovskite (DP) materials has surged in recent decades owing to the strong coupling among charge, spin, orbital, and lattice degrees of freedom, which gives rise to a wide range of emergent physical phenomena, including multiferroicity \cite{multiferroic1,multiferroic2,multiferroic3,multiferroic4,multiferroic5}, large exchange bias \cite{eb1,eb2,eb3,eb4,eb5,eb6,eb8,macchiutti2024,sahoo2021,cmo2,cmo3,cmo4,asd_x1,hole1}, and colossal magnetoresistance \cite{cmr1,cmr2,cmr3}. Owing to these properties, DP materials have been extensively explored as functional platforms for diverse technological applications, including spintronic and magnetic memory devices \cite{Tang2022, Mustafa2022, Kumari2024} cryogenic magnetic refrigeration \cite{eb7,mce1,mce2,mce3,mce4}, renewable energy generation \cite{reg1,reg2,reg3,ecpc3,Lei2021}, photovoltaics and optoelectronics \cite{Roknuzzaman2019,Luo2018,Zhao2018,Lei2021}. A $B$-site cationic ordering exists in $A_{2}BB’O_{6}$ [$A$: rare/alkaline-earth atoms, $B, B'$: transition metal atoms] DP oxide structure, where the $BO_{6}$ and $B'O_{6}$ octahedra [\autoref{Fig. (1)}(d)] are alternately placed along the three crystallographic directions, as shown in \autoref{Fig. (1)}(a), which results in a doubling of the unit cell as compared to a single perovskite structure, which does not have a $B$-site cation ordering [\autoref{Fig. (1)}(b)]. Large differences in ionic radii and cationic charge between the $B$ and $B'$ cations seem to be necessary to stabilize a double perovskite structure \cite{galasso1959,vasala2015,king2010}. The titular double-perovskite compound Nd$_{2}$CoMnO$_{6}$ is one such example, where significant charge difference (Co is in $+2$ state and Mn is in $+4$ state) as well as large size difference (ionic radii of six-coordinated Co$^{2+}$ and Mn$^{4+}$ions are $0.745$ \AA \, and $0.53$ \AA, respectively \cite{shannon_nd}) exists for the participating B-site cations. Due to the ferromagnetic Co$^{2+}-$ O$^{2-}-$Mn$^{4+}$ $180^\circ
$ super-exchange interaction, double perovskite Nd$_{2}$CoMnO$_{6}$ orders ferromagnetically at $\sim 160$ K. When such a criteria are not met in case of the participating $B$ and $B'$ cations, inherent B-site cationic disorder is introduced, i.e. the alternate placement of $BO_{6}$ and $B'O_{6}$ octahedra gets disturbed which is referred to as antisite disorder (ASD). In the context of magnetic superexchange interactions, ASD results in the introduction of additional antisite couplings, namely $B$-O-$B$ or $B'$-O-$B'$, in place of the spontaneous $B$-O-$B'$ couplings of the ordered structure. Also, in presence of heterovalent cation doping, like for Sr$^{2+}$ ion doping at the Nd$^{3+}$ ionic site in the present case, new charge-states for the participating B-cations get introduced to maintain electrical charge neutrality of the compound. This leads to many emergent magnetic super-exchange interactions, as for example, introduction of Co$^{3+}-$ O$^{2-}-$Mn$^{4+}$, Mn$^{2+}-$ O$^{2-}-$Co$^{3+}$ etc., which were absent in the undoped compound and many of which are inherently conflicting in nature. Further introduction of multiple valence states for the B-site cations in presence of heterovalent doping can lead to a compromise in the required charge-difference and size-difference for the participating B-cations for a double perovskite compound and can introduce ASD. Presence of such varied and often conflicting (i.e. presence of both ferromagnetic [FM] and antiferromagnetic [AFM]) super-exchange interactions driven by heterovalent doping as well as ASD can cause broadening of the magnetic transitions and eventually leads to the emergence of glassy spin-dynamics for the cation-disordered compound with increasing doping percentage. Further, due to the unavoidable spatial inhomogeneity associated with the distribution of the dopant ions, the sample gets inevitably phase separated into regions of more cation-ordered FM regions and cation-disordered AFM/magnetically-disordered/glassy magnetic regions.\par

\begin{figure*}
\scalebox{1.0}
{\includegraphics[width=\textwidth]{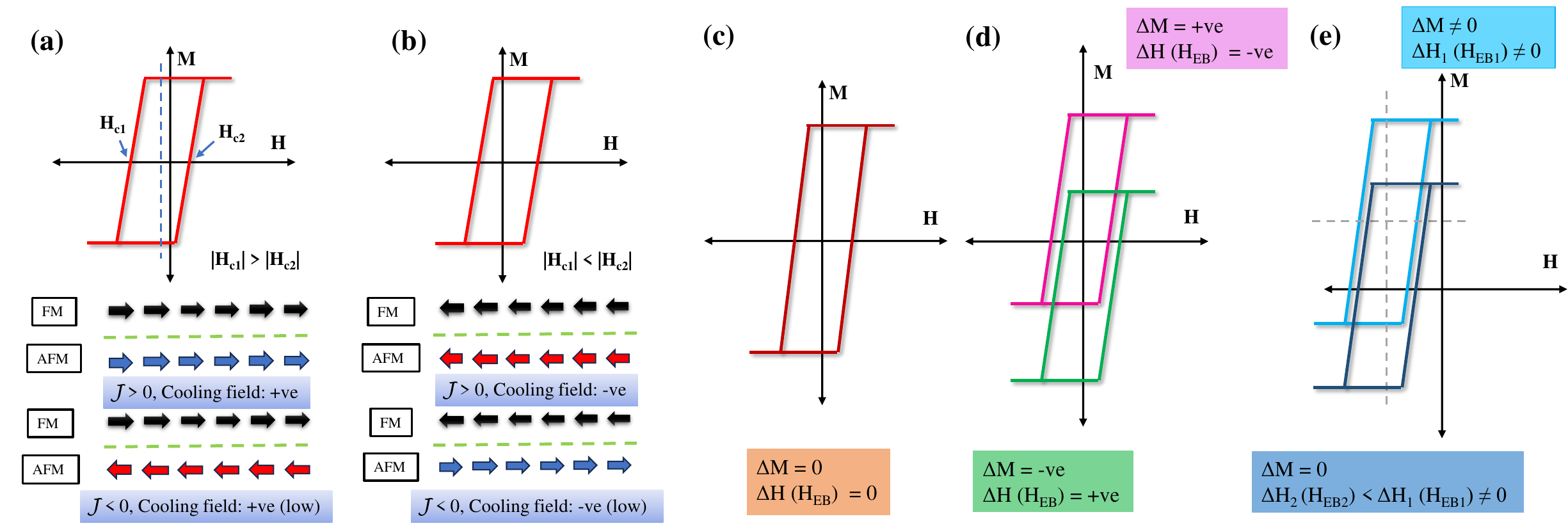}}
\caption{(color online) (a) and (b) show various scenarios of ferromagnetic--antiferromagnetic interfaces subjected to different cooling fields and interface couplings, resulting in exchange bias (EB) phenomena. 
(c) shows a perfectly symmetric magnetization loop with its centroid at $(0,0)$ and having $0$ EB. 
(d) shows vertically shifted versions of (c); note that the apparent asymmetry along the applied field axis ($H$) gives rise to ``exchange-bias-like'' behavior, although this is merely a manifestation of a minor loop and not intrinsic EB. 
(e) shows the most common type of loop encountered in real magnetic systems (sky-blue loop with a vertical shift). This loop has been shifted downward to compensate for the minor-loop contribution, yielding the navy-blue loop with zero vertical shift and smaller EB than the vertically shifted sky-blue loop.
}\label{Fig. (2)}
\end{figure*}

Interestingly, the presence of interface anisotropy at the boundary between co-existing FM and AFM phases can lead to the emergence of the Exchange Bias (EB) effect \cite{ebg1,ebg2,ebg3,ebg4,mb1}. In addition to the interface between FM and AFM materials, the EB effect has also been documented in interfaces involving FM/ferrimagnetic (FiM), FM/Spin-Glass (SG), FM/Cluster-Glass (CG), AFM/SG and so on. The utilisation of EB materials has resulted in the development of several applications, including magnetic recording, spin valve readback heads, MRAM memory circuits, permanent magnets, magnetic sensors and other spintronic devices \cite{eba1,eba2,eba3,eba4,eba5,eba6}. Currently, materials possessing a significant EB and coercive field are subjects of immense scientific and technological investigations. EB occurs when the hard magnetization behavior of antiferromagnetic layer produces an interface bias magnetic field to influence the soft magnetization curve of the overlapping ferromagnetic region \cite{ebg2}. The result is a horizontal shift in the ferromagnet’s magnetization (M) vs. magnetic field (H) curve in the opposite direction of the bias magnetic field (\textit{ideally}, with no vertical shift i.e. shift along the magnetization axis for fully compensated AFM layers and extremely small vertical shift for uncompensated AFM layers). For example, if the M(H) hysteresis loop cuts the M $= 0$ axis at H$_{c1}$ and H$_{c2}$ on the descending and ascending sides of H axis, respectively, the EB can be defined on the basis of exchange field H$_{EB}$ = (H$_{c1} + $H$_{c2}$)/$2$ [as shown in \autoref{Fig. (2)}(a)]. The amplitude and sign of EB are heavily influenced by the direction and strength of the bias field, as well as the nature and anisotropies of the interface coupling, as has been schematically shown in \autoref{Fig. (2)}(a) and \autoref{Fig. (2)}(b). Normally, the shift of the M(H) hysteresis loop happens in the opposite direction to that of applied cooling field (which is known as the normal exchange bias or NEB) [\autoref{Fig. (2)}(a) and \autoref{Fig. (2)}(b)]. But in some rare cases the shift appears in the direction of cooling field, which is an unusual phenomenon and is known as the inverse or positive exchange bias (IEB or PEB). The origin of the IEB phenomena remains poorly understood till date backed with only a handful of examples \cite{eb3,eb5,eb6,eb9,peb1,peb3,peb4,peb5,peb6,peb7,salamon1,sgieb4}.

One commonly overlooked aspect in the investigation of EB is the adequate verification of magnetic saturation for the specific compound being studied. The phenomenon of non saturation, often referred to as the minor loop, is characterized by a notable displacement along both the field axis and magnetization axis, combined with some asymmetry observed in the M(H) hysteresis loop \cite{harres2016}. Such examples of minor loops are schematically shown in \autoref{Fig. (2)}(d) (which are vertically shifted versions of \autoref{Fig. (2)}(c)), where a shift along the magnetization axis (M) automatically produces an apparent displacement along the field axis (H), which is not a signature of true EB. In real magnetic systems possessing some degree of intrinsic EB, one is more likely to encounter a hysteresis loop similar to the vertically shifted example in \autoref{Fig. (2)}(e) (in sky blue color), where the vertical shift arising from the minor loop introduces artifacts into the intrinsic EB. The degree of shift and/or asymmetry is contingent upon the level of magnetic saturation that has been attained. In systems where the magnetic anisotropy field (the field necessary to achieve magnetic saturation) is notably greater than what can be attained in routine laboratory measurement systems (which is particularly true for many DP systems), there exists a significant amount of research that erroneously interprets the minor loop phenomenon as an intrinsic EB effect and this stream of research continues to expand at a rapid pace \cite{geshev2009,geshev2008,geshev2008b,harres2015,klein2006}. In this regard, we have recently devised a methodology, which has been demonstrated to mitigate such artifacts arising from magnetic non-saturation effects and be able to extract the intrinsic EB effect in such cases \cite{mishra}. Schematically, this is shown in \autoref{Fig. (2)}(e), where the vertically shifted loop (sky blue color) has been translated downward to eliminate the vertical shift (now in navy blue color), thereby eliminating the contribution of spins that don't rotate reversibly during the M(H) loop field cycling and yielding the intrinsic EB. Following preceding discussion, it is clear that heterovalent ion-doped double perovskite oxides provide an ideal playground to host varied kinds of magnetic interfaces between magnetically phase separated FM and AFM/magnetic-disordered/glassy regions and can potentially harbour exotic and large EB effects. In addition, EB effects in such heterovalent-ion-doped double perovskite oxides are expected to be tunable through variations in the extent of ASD, the presence of multiple cation valence states, and ionic-size-driven redistribution of magnetic superexchange interactions. Notably, large EB fields ($\sim 3$ kOe at $T = 5$ K) have been reported in $RE_{2-x}$A$_{x}$CoMnO$_{6}$ compounds (where $RE$ = La, Y and A = Ca, Sr, Ba) for a doped specific composition, i.e. $RE$ = La, A = Sr and $x = 0.5$ \cite{eb8}. Interestingly, the $x$-dependence of EB properties in La$_{2-x}$A$_{x}$CoMnO$_{6}$ ($A$ = Sr, Ca) for ($0 \leq x \leq 1$) gave conflicting results, with some reports suggesting a non-monotonic dependence of EB on $x$ \cite{cmo2,asd_x1,sahoo2021}, while others report a monotonic trend for EB field on $x$ for the similar family of compounds \cite{hole1}. Such inconsistencies in EB results seem to have stem from minor-loop effects in the measured M-H loops which mask the intrinsic EB properties, and, thus a systematic $x$-dependent study on the inter-relation between structural, electronic, magnetic and intrinsic EB properties of a $RE_{2-x}$A$_{x}$CoMnO$_{6}$ series is absolutely necessary and crucial to elucidate the true nature and potential of this promising family of compounds.

Keeping these points in mind, we have synthesized polycrystalline bulk samples of Nd$_{2-x}$Sr$_{x}$CoMnO$_{6-\delta}$ ($0 \leq x \leq 1$) series, and conducted a comprehensive analysis of its structural and magnetic characteristics, with a particular emphasis on EB. Using X-ray and neutron powder diffraction, we have succesfully characterized the increasing disorder in the series with hole doping. X-ray absorption spectroscopy reveals gradual hole transfer to Co$^{2+}$ with doping, converting it to Co$^{3+}$, while Mn valence stays unchanged. Combining DC and AC magnetization investigations, we elucidate a rich magnetic phase diagram of this series. Seemingly the cation-ordered region of the sample continue to remain ferromagnetic with  increasing $x$, whereas the cation-disordered region (predominantly involving other cationic valencies) evolves gradually (with increasing $x$) from AFM to a magnetically disordered region. Around $x = 0.625$, the magnetically disordered region evolves into a magnetic glassy state. The low-tempearture magnetic hysteresis loops exhibit significant horizontal as well vertical shift ($\sim 14\%$ of average saturation magnetization for $x = 0.75$!), a clear signature of strong minor loop presence. Such effects have been effectively mitigated through the adjustment of the vertical magnetization shift to zero \cite{mishra}, and thereby intrinsic EB field have been evaluated for each composition. Our findings elucidate the presence of IEB in disguise of NEB for $x = 0.75$, which shows robust IEB even in a high cooling field of $6$ Tesla. These EB results are deeply concomitant with the presence of frustrated spins, thereby highlighting the pertinent role of interfaces involving magnetically glassy regions in the context of EB in DP systems. 

\section{Experimental Details and Methodologies}
The Nd$_{2-x}$Sr$_{x}$CoMnO$_{6-\delta}$ $(0 \leq x \leq 1)$ samples (in short, NSCMO) were synthesized by the conventional solid-state reaction method. Appropriate amounts of Nd$_2$O$_{3}$ (Alfa Aesar, $99.99\%$), SrCO$_{3}$ (Alfa Aesar, $99.99\%$), CoO (Alfa Aesar, $99.995\%$) and MnO$_{2}$ (Alfa Aesar, $99.996\%$) powders were mixed and ground in a mortar until a fine, homogeneous mixture was obtained. Nd$_2$O$_{3}$ was pre-heated at $900^{\circ}$ to remove absorbed moisture and gas. The mixture was first heated at $1200^{\circ}$ for $24$ hours, followed by an intermediate grinding and reheating at $1300^{\circ}$ for another $24$ hours. Slow cooling rate of $1^{\circ}$C/min was adopted to ensure maximum ordering of $B$ site cations. The crystallographic details of the polycrystalline bulk powder sample have been verified by a divergent beam laboratory X-ray diffraction (Empyrean, PANalytical) apparatus with Cu $K{_\alpha}$ source ($\lambda$ = $1.5405$ \r{A}) at room temperature. Further crystallographic and magnetic structure analysis were performed using the Neutron Powder Diffraction ($\lambda$ = $1.667$ \r{A}) facility of the \textsc{Pearl Diffractometer} at the TU Delft Reactor Institute. The similar scattering factors of Co and Mn ions complicate the identification of Co/Mn ordering solely through X-ray diffraction (XRD) patterns. On the other hand, Neutron powder diffraction (NPD) experiments can identify Co, Mn cation ordering due to a significant difference in their neutron scattering lengths (Mn $\sim -3.73$ fm and Co $\sim 2.49$ fm) \cite{scattering}. Consequently, both techniques have been employed to extract the structural data.

Rietveld refinements \cite{rietveld1969} were performed using \textsc{FullProf Suite} software \cite{rodriguez1993} and crystal structures were generated using \textsc{Vesta} \cite{vesta2011}. Elemental compositions and their homogeneity were reconfirmed by using multiple field emission gun scanning electron microscopes (Merlin FEGSEM by \textsc{Zeiss} with Gemini II column and JEOL Ultra-High Resolution FESEM) (elemental composition results are shown in Appendix~\ref{app:edax} of the Supplemental Material [SI]). The electronic valence states of Co and Mn have been verified using X-ray Absorption Spectroscopy facility of the Variable Polarization XUV beamline P04 at PETRA III, Deutsches Elektronen-Synchrotron (DESY) in Total Electron Yield (TEY) mode. XMCD signal were obtained using the Max-P04 end-station of P04 at PETRA III at $110$ K temperature and in presence of a $0.8$ Tesla magnetic field using a liquid $N_2$ flow-cryostat. In addition, X-ray photoelectron spectroscopy (PHI $5000$ Versa Probe III Scanning) equipped with an monochromatic Al $K_{\alpha}$ source ($h\nu$ = $1486.6$ eV) at room temperature was used to probe the Nd and Sr-valence states. The samples were sputtered using Ar ions of energy $2$ keV for sufficient time before recording the spectra. Also, Raman spectrocopic measurements were done using a LabRAM HR Evolution Raman spectrometer by Horiba Scientific using a $532$ nm laser source to further investigate the structural disorder.

The temperature and field-dependent magnetization measurements have been carried out in a magnetic property measurement system (MPMS3, Quantum Design) in the temperature range of $8$ K to $300$ K. To ensure complete demagnetization and thermal stabilization of the samples, the samples were first heated to the paramagnetic (PM) state (room temperature), then cooled to the desired tempearature followed by a wait time of two-minutes before starting a new measurement. Certain well-defined protocols were followed to keep the trapped field of the superconducting magnet minimal and same for all the compositions \cite{mishra}. For AC susceptibility measurements, the samples underwent cooling to the minimum achievable temperature value from the paramagnetic (room temperature) state, while being subjected to a ``zero magnetic field'' (zero-field-cooled). Readings were taken during the warm-up cycle at a very slow temperature ramp rate while subjecting the sample to an AC field of $4$ Oe and zero DC field. Magnetization vs. field measurements were performed in both zero-field-cooled (ZFC) and field-cooled (FC) mode at the maximum available sweep field of $7$ Tesla and at $8$ K (with a uniform field step size of $50$ Oe). In the context of exchange bias, the ZFC mode is known as the spontaneous exchange bias (SEB) and likewise, the FC mode is known as the conventional exchange bias (CEB). For each of the ZFC and FC mode, two separate protocols of field sweeping were followed: P mode ($0$ $\to$ $+7$ T $\to$ $-7$ T $\to$ $+7$ T $\to$ $0$) and N mode ($0$ $\to$ $-7$ T $\to$ $+7$ T $\to$ $-7$ T $\to$ $0$), to check for complete reversibility of EB. In addition, iodometric titration \cite{titration1,titration2,titration3,titration4} were performed on several compositions to quantify the presence of oxygen vacancies: accurately weighed samples ($\sim 20$ mg) were dissolved in $8$ N HCl and excess ($\sim 10 \%$) KI; high-valent Co and Mn ions quantitatively oxidise I$^{-}$ to I$_2$ while being reduced to $+2$. The liberated iodine was back-titrated with standardized $\sim$ 0.1~N \ce{Na2S2O3} using starch as indicator, and oxygen-vacancy ($\delta$) was calculated from the consumed thiosulfate equivalents via the charge-neutrality-derived relation shown in the Appendix~\ref{app:iodometric} of SI \cite{SI}. \par
The electronic structure of $x = 0$ was investigated using density functional theory (DFT) in the Vienna Ab initio Simulation Package (VASP) to understand the hole and electron absorbing behavior of transition metal and rare-earth atoms \cite{kresse1996}. The projector-augmented wave (PAW) method was employed with the Perdew-Burke-Ernzerhof (PBE) functional in the generalized gradient approximation (GGA) \cite{perdew1996}. Strong correlations in Nd $4f$, Co $3d$, and Mn $3d$ states were treated via DFT+$U$ (Dudarev formulation \cite{dudarev1998}) with $U=6.0$, $3.0$, and $4.0$ eV, respectively ($J=0$ eV). Structural relaxation (ionic positions and cell parameters) started from an antiferromagnetic configuration with spin polarization and initial moments of $-1~\mu_\text{B}$ (Nd), $+4~\mu_\text{B}$ (Co/Mn), and $0~\mu_\text{B}$ (O). A 9$\times$9$\times$6 Monkhorst--Pack $k$-mesh was used for Brillouin zone sampling. Convergence thresholds were 10$^{-6}$ eV for electronic self-consistency and $-0.01$ eV/$\AA$ for ionic forces. Finally, Density of states (DOS) were calculated from the converged charge density in a static run with projected orbitals on a 1000-point grid with the same set of $k$-mesh. 

\section{Results and Discussions}
\subsection{Crystal structure}
\begin{figure*}
\scalebox{0.9}
{\includegraphics[width=\textwidth]{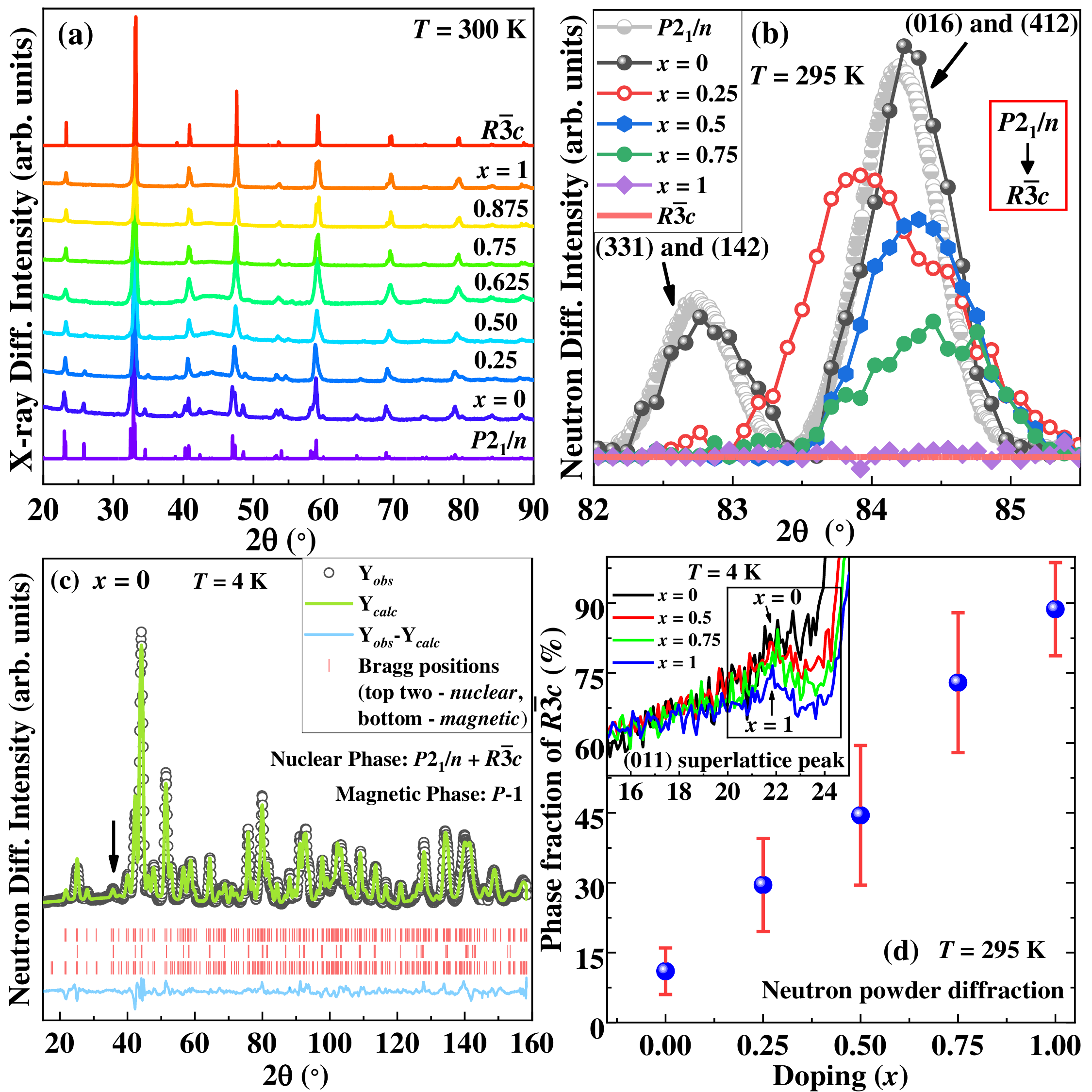}}
\caption{(color online)  (a) Room temperature powder X-ray Diffraction (XRD) pattern of the NSCMO series. The references were taken from ICSD database. (b) Gradual evolution of phase from $P2_1/n$ to $R\overline{3}c$ with hole doping as seen in the room temperature neutron powder diffraction (RT-NPD) data. All the reflections $(331), (142), (016),$ and $(412)$ are forbidden in $R\overline{3}c$ due to the rhombohedral centering condition: $ -h + k + l = 3n$, ruling out $Pnma$ or $Pbnm$ (where only $(016)$ is forbidden but others are allowed). (c) Rietveld refinemnt of $4$ K NPD data of $x = 0$. The most prominent commensurate magnetic contribution has been highlighted by a downward arrow. (d) Fraction of disordered $R\overline{3}c$ phase as obtained from the Rietveld refinements of RT-NPD data. Inset shows the comparison of $(011)$ superlattice peak in the $4$ K NPD data for different $x$ compositions. The relative weakening of intensity for higher $x$ members indicate increased disorder with doping.}\label{Fig. (3)}
\end{figure*}

\begin{figure*}
\scalebox{0.9}
{\includegraphics[width=\textwidth]{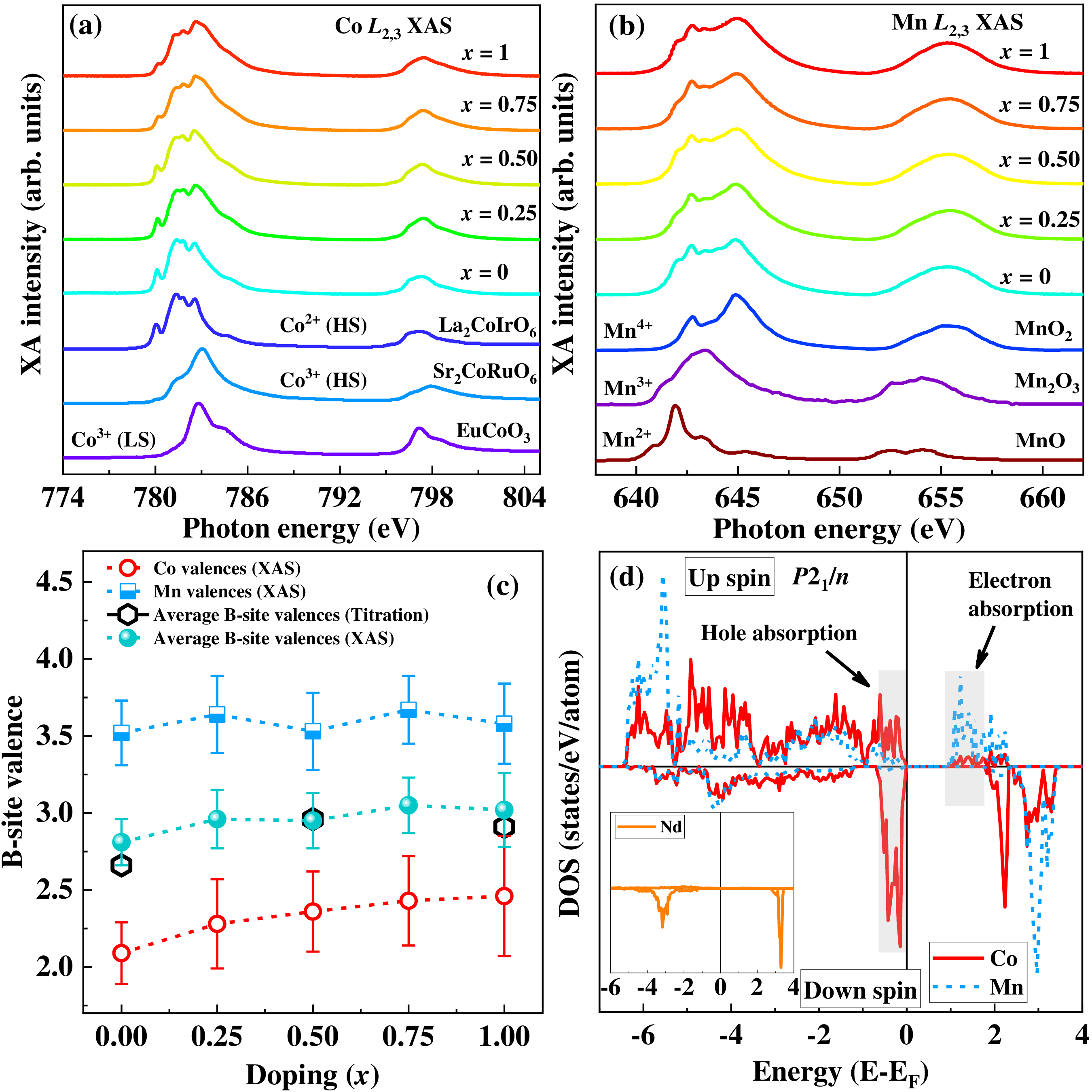}}
\caption{(color online) (a) Co $L_{2,3}$ and (b) Mn $L_{2,3}$ edge XAS spectra along with the reference spectra. (c) Effective Co and Mn valence as well as average B-site valence as a function of $x$ as determined from the weighted linear convolution of the reference spectra. These numbers are compared with the average B-site valence obtained from iodometric titration. (d) Density of states (DOS) plotted for a B-site ordered structure of $x = 0$ taking Fermi energy level as $0$. Hole absorbing behavior of Co and electron absorbing behavior of Mn have been highlighted. Inset shows Nd states deep below the Fermi energy.}\label{Fig. (4)}
\end{figure*}

It is important to note that $T$-dependent NPD results clearly rule out any temperature-dependent structural phase transition down to $4$ K for all $x$ compositions (representative T-dependent NPD data are shown for $x = 0$ and $x = 0.5$ in Fig.~\ref{fig:S1} of SI \cite{SI}). To investigate how the B-site cation ordering evolves with increasing $x$, we first discuss the room-temperature NPD results. We focus on a narrow $2\theta$ window between $2\theta = 82^{\text{o}}$ to $85^{\text{o}}$ (as seen in \autoref{Fig. (3)}(b)), which captures the characteristic NPD peaks for both the B-site cation ordered $P2_{1}/n$ phase as well as the B-site cation disordered $R\overline{3}c$ phase. The peaks corresponding to the reflection planes $(331), (142), (016)$ and $(412)$, which appear in this $2\theta$ window, are allowed for the $P2_{1}/n$ phase but are forbidden for the $R\overline{3}c$ phase due to the rhombohedral centering condition: $ -h + k + l = 3n$. Interestingly, the $(016)$ peak, which is quite strong for the $x = 0$ composition, is a unique signature peak for the $P2_{1}/n$  space group and is forbidden both in $R\overline{3}c$ as well as for other possible B-site disordered phases, like orthorhombic $Pbnm$ or $Pnma$. These results clearly elucidate that $x = 0$ composition stabilizes primarily in the cation ordered $P2_{1}/n$ space group. The same is also evident from the characteristic presence of the superlattice NPD peak (arising from B-site cation ordering), as seen in the $4$ K NPD data [inset of \autoref{Fig. (3)}(d)] as well as room temperature (RT) NPD data in Fig.~\ref{fig:S2} of SI \cite{SI}. For the $x = 1$, all the $(331), (142), (016)$ and $(412)$ peaks are nearly absent as seen in \autoref{Fig. (3)}(b). Since the $(331), (142)$ and $(412)$ NPD peaks are allowed for cation-disordered $Pbnm/Pnma$ phases, the absence of them for $x = 1$ clearly suggest that there is a gradual evolution of the structural phase from primarily $P2_{1}/n$ for $x = 0$ to predominantly $R\overline{3}c$ for $x = 1$. The NPD results are also in consistence with the RT XRD results, as shown in \autoref{Fig. (3)}(a), which also suggest a gradual evolution of the structural phase from nearly $P2_{1}/n$ for $x = 0$ to nearly $R\overline{3}c$ for $x = 1$. Importantly, no impurity phase's peak were observed within the instrument's resolution limit for any of the $x$ compositions, as evidenced from \autoref{Fig. (3)}(a) as well as Fig.~\ref{fig:S3} of SI \cite{SI}. Accordingly, the room temperature NPD data spectra were fitted (by employing Rietveld refinement using the \textsc{FullProf Suite} software) by considering a combination of both B-site cation ordered $P2_{1}/n$ and B-site cation disoredred $R\overline{3}c$ space groups for all $x$ compositions. The obtained refinement results elucidate a near linear evolution from $90 \%$ $P2_{1}/n$ phase for $x = 0$ to nearly $90 \%$ $R\overline{3}c$ phase for $x = 1$ composition, as shown in \autoref{Fig. (3)}(d). The evolution from B-site cation ordered structure for $x = 0$ towards B-site cation disordered structure of $x = 1$ can also be qualitatively captured using room-temperature Raman spectroscopy. It is well documented that stretching (S) and anti-stretching (AS) vibration modes of the $BO_{6}$ octahedra become broader and of weaker intensities in the B-site cation disordered phase as compared to the B-site cation ordered phase \cite{raman1,raman2,raman3,raman4}. Indeed, the S and AS Raman peaks (shown in Fig.~\ref{fig:S4} of SI \cite{SI}), which are very sharp and of considerable intensity in $x = 0$, become broader and of progressively weaker intensity with increasing proportion of $x$ in Nd$_{2-x}$Sr$_{x}$CoMnO$_{6}$, thus being in qualitative consistency with NPD and XRD results. 

\subsection{Valence states}

Next we investigate the evolution of the valence states of the constituent ions to understand the origin of increase of the cation-disordered $R\overline{3}c$ phase fraction with progressively increasing Sr (i.e. hole) doping. Nd and Sr cations are found to be in the expected $+3$ and $+2$ stable valence states, respectively, throughout the series as elucidated through X-ray Photoelectron Spectroscopy (XPS) investigations (representative XPS spectra are shown in Fig.~\ref{fig:S5} of SI \cite{SI}). However, to investigate the valence state of the transition metal cations, namely Co and Mn, which can exhibit multiple valence states, we resort to the more bulk-sensitive, characteristic electronic structure probe, i.e. X-ray Absorption Spectroscopy (XAS), which probes the bulk electronic structure as compared to XPS, which detects mostly the surface electronic structure. We first discuss the evolution of the Mn valence states. As seen from \autoref{Fig. (4)}(b), the Mn $L_{2,3}$ XA spectra of the various $x$ compositions appear broadly similar suggesting that the average Mn valency remain nearly similar across the series. When compared with the Mn $L_{2,3}$ XA spectra of the reference compounds \cite{MnXAS,MnXAS1} which contain only one of Mn valence state, i.e. either $+4$, $+3$ or $+2$ in a similar octahedral environment, it becomes apparent that the Mn ions in the present series of compounds are mostly in the $+4$ valence state and with smaller contributions from $+3$ and $+2$ valence states. We next perform a quantitative analysis by trying to reproduce the Mn $L_{2,3}$ XA experimental spectra for various compositions as a linear sum of various reference XA spectra for different Mn valence states. Such an analyses provide us with a quantitative estimate of the relative presence of various Mn valence states in different $x$ members, as shown in Table~\ref{tab:Mn_XAS} of SI \cite{SI}. An important quantity, i.e. the average Mn valency, can also be extracted from such an analysis, is seen to be remain nearly constant around $3.6$ for all $x$ [\autoref{Fig. (4)}(c)]. \par
While the Mn ions remain in their high-spin state in oxides [\autoref{Fig. (1)}(g) and \autoref{Fig. (1)}(h)], Co ions in oxides can occur in multiple valence as well as spin states [\autoref{Fig. (1)}(c), \autoref{Fig. (1)}(e) and \autoref{Fig. (1)}(f)] \cite{zobel2002,podlesnyak2006,fauth2001,HSIS1,HSIS2,HSIS3,HSIS4,HSIS5,HSIS6,HSIS7,HSIS7,HSIS8,HSIS9}. We first discuss the valence state of Co in $x = 0$ compound. On comparing with Co $L_{2,3}$ XA spectra with the reference compounds \cite{iridium} which contain Co in a specific valence as well as spin state, as shown in \autoref{Fig. (4)}(a), it becomes apparant that Co valence state in $x = 0$ is nearly $+2$ with small contribution of Co$^{3+}$ in high-spin (HS) state. However, the Co $L_{2,3}$ XA spectra shows continuous spectral shape evolution with increasing $x$ (i.e. hole doping). On performing a similar analysis of reproducing the experimental spectrum as a linear sum of various reference Co spectra, we obtain the relative contributions of various valence and spin states of Co in all compositions across the series. Unlike in case of Mn ions, the average valence of Co ion increases steadily with increasing $x$ (as shown in Table~\ref{tab:Co_XAS} of SI \cite{SI} and \autoref{Fig. (4)}(c)), thereby, elucidating that the doped holes from Sr doping reside mostly on Co site. The presence of Mn in lower oxidation states than $+4$, i.e. in the $+3$ and $+2$ valence states, seems to be driven by the localization of the doped electrons arising from the emergence of oxygen vacancies. To investigate the origin of the doped holes (through Sr$^{2+}$ doping for Nd ions) residing mainly on the Co ions and the doped electrons (from the presence of oxygen vacancies) primarily localizing on the Mn ions, we have computed the density-of-states (DOS) of Nd$_{2}$CoMnO$_{6}$ using first-principles based density functional theory. Indeed, as seen through \autoref{Fig. (4)}(d), the states close to the valence band maximum are dominated by Co states, where the doped holes get localized. Similarly, the states near the conduction band minimum, where doped electrons localize, are dominated by the Mn states. The oxygen contents of these samples were independently determined through iodometric titration, as detailed in the Appendix~\ref{app:iodometric} of SI \cite{SI}. The oxygen contents (and corresponding average B-site cation valences) determined by iodometric titration are in excellent agreement with the values obtained using XAS (see Table~\ref{tab:delta} and Table~\ref{tab:B_valence} in the SI). The steady increase of the Co$^{3+}$ (HS) and Co$^{3+}$ (LS) contents with increasing $x$, both of which have a lower charge and ionic size (ionic radii of octahedrally coordinated Co$^{3+}$ (HS) and Co$^{3+}$ (LS) are $0.61$ \AA \, and $0.545$ \AA, respectively \cite{shannon_nd}), as compared to Co$^{2+}$ (HS) (ionic size of octahedrally coordinated Co$^{2+}$ (HS) is $0.745$ \AA \, \cite{shannon_nd}) indeed helps to understand the driving force for steady increase of the B-site cation-disordered $R\overline{3}c$ phase fraction with increasing $x$, as seen in \autoref{Fig. (3)}(d).

\subsection{Magnetic properties}

\begin{figure}
\scalebox{0.5}
{\includegraphics[width=\textwidth]{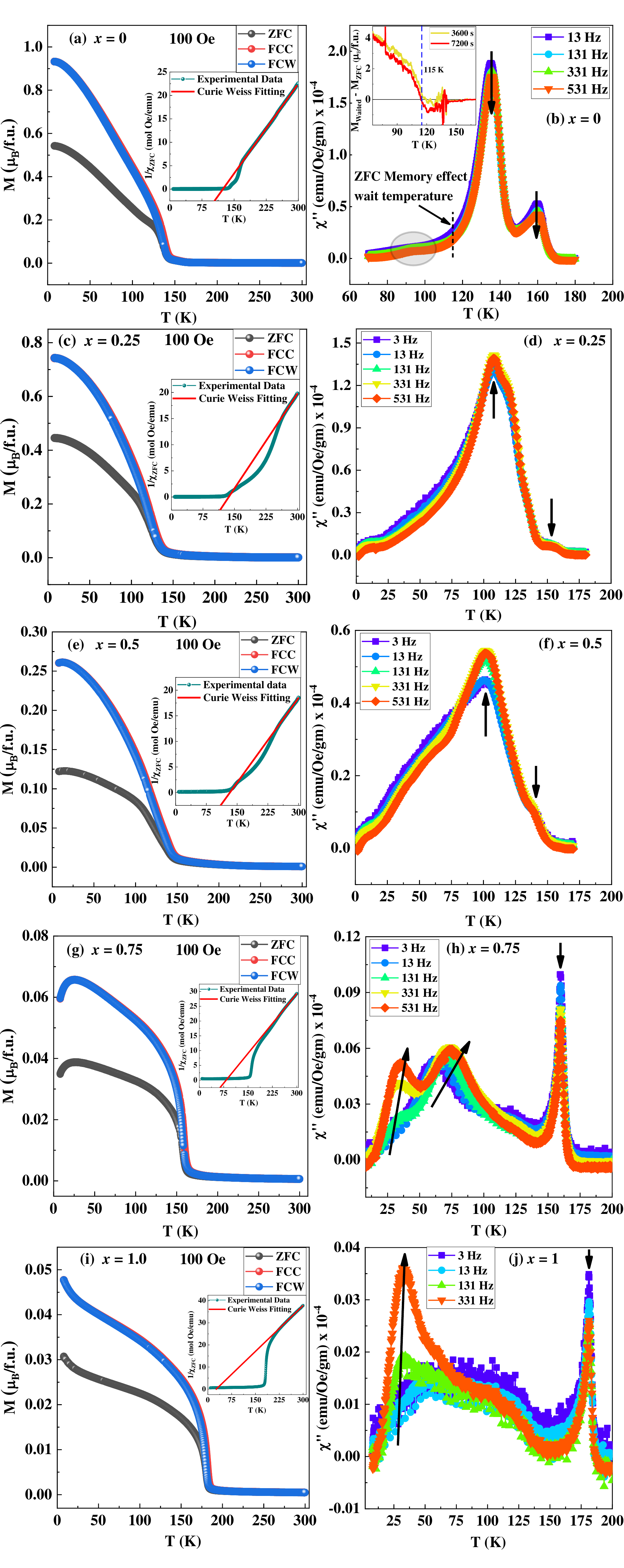}}
\caption{(color online) (a), (c), (e), (g) and (i) $100$ Oe DC magnetization data for various $x$ compositions; and (b), (d), (f), (h) and (j) the corresponding AC susceptibility data (imaginary part). Insets show $1/\chi$ plot and standard Curie-Weiss linear fitting of the ZFC magnetization data. The vertical arrows in the AC data indicate long-range transitions, while the skewed arrows indicate glassy transitions.}\label{Fig. (5)}
\end{figure}

Presence of both the cation ordered $P2_{1}/n$ phase containing Co$^{2+}$/Mn$^{4+}$ ions and the cation disordered $R\overline{3}c$ phase containing mixtures of Co$^{3+}$/Co$^{2+}$/Mn$^{3+}$/Mn$^{4+}$/Mn$^{2+}$ ions in the samples are expected to significantly affect the magnetic properties. The $100$ Oe M(T) data for the parent composition $x = 0$ are shown in \autoref{Fig. (5)}(a). Consistent with the presence of above phases, the $x = 0$ sample is found to exhibit two ferromagnetic transitions around $\sim 160$ K (referred as T$_{c1}$) and $\sim 136$ K (referred as T$_{c2}$), which are understandably driven by the B-site cations since the rare-earth (Nd ions) lattice orders independently at much lower temperatures \cite{lynn1990,tranquada1996}. The above observations are consistent with the literature for $x = 0$ \cite{ncmo1}, where, like other $RE_{2}$CoMnO$_{6}$ DP systems, the T$_{c1}$ and T$_{c2}$ transitions have been reported to be driven by the ferromagnetic Co$^{2+}-$ O$^{2-}-$Mn$^{4+}$ and Co$^{3+}-$ O$^{2-}-$Mn$^{3+}$/Co$^{3+}-$ O$^{2-}-$Mn$^{4+}$ superexchange interactions, respectively \cite{eb3,eb8,macchiutti2024,cmo1,cmo2,cmo3,cmo4,asd_x1,ncmo1,ncmo2,ncmo3,goodenough1,co3mn4_1}. Thus, the T$_{c1}$ transition can be assigned to the $P2_{1}/n$ phase and the T$_{c2}$ transition to the $R\overline{3}c$ phase in our samples. Indeed, the XMCD results on $x = 0$ at $110$ K (which is below T$_{c1}$ and T$_{c2}$) indicate a parallel spin alignment for Co and Mn as shown in Fig.~\ref{fig:S6} of SI \cite{SI}, in consistence with it's ferromagnetic ordering. The M-H loop of $x = 0$ at $8$ K exhibits larger coercive field of $\sim 11.5$ kOe and saturation moment of $\sim 6.90$ $\mu_\text{B}$/f.u. at $7$ Tesla field, which are also consistent with the earlier report on well-ordered Nd$_{2}$CoMnO$_{6}$ sample \cite{ncmo1} and indicates high degree of B-site cation ordering, as also verified by our diffraction results. Similarly, the $4$ K neutron diffraction spectrum of $x = 0$, which captures it's magnetic structure, can also be described well (as shown in \autoref{Fig. (3)}(c)) considering a ferromagnetic alignment between the Co and Mn spins with estimated local moment values for Co/Mn and Nd as $3.776 \, \, \mu_\text{B}$/atom and $-0.5 \, \, \mu_\text{B}$/atom, respectively, which are in good agreement with the earlier report \cite{ncmo3}. It is important to note that, the most prominent magnetic peak emerges around $36^{\text{o}}$ (as shown in \autoref{Fig. (3)}(c) by a downward arrow) and the progressive manifestation of this peak with temperature is shown in the inset of Fig.~\ref{fig:S1}(a) of SI \cite{SI}. For the other $x$ members, this magnetic peak gets strongly overlapped with an emerging nuclear peak around the same position [shown in the inset of Fig.~\ref{fig:S1}(b) of SI \cite{SI} for $x = 0.5$], making it difficult to extract the magnetic structure as well as the respective magnetic moments of the constituent cations. 

Now, to rule out any glassy dynamics associated with these transitions, we have performed dc memory effect experiment \cite{md1,memory1,memory2} at $115$ K temperature (well below the prominent transitions) and found no dip in the subtracted magnetization data [inset of \autoref{Fig. (5)}(b)]. It is noteworthy to point out that earlier reported result by Sazonov \textit{et al.} \cite{ncmo3} presents a spin-glass like phase around $\sim 90$ K, attributed to the presence of inherent ASD. In our sample, although a feeble feature can be seen around $\sim 90$ K in the AC $\chi^{''}$ data (highlighted in \autoref{Fig. (5)}(b) with a grey oval), we could not observe any conclusive frequency-dispersion, thereby we rule out any glassiness in our $x = 0$. \par Now, the two magnetic transitions corresponding to T$_{c1}$ and T$_{c2}$ exhibit qualitatively different evolution with increasing $x$ as seen in \autoref{Fig. (5)} and Fig.~\ref{fig:S7} of SI \cite{SI}. The magnetic transition at T$_{c1}$, although becomes slightly broader between $0$ to $0.5$ and thereby recovers sharpness for higher $x$ members, occur nearly around the same temperature window for all $x$ members. In sharp contrast, the magnetic transition corresponding to T$_{c2}$ becomes significantly broader, weaker and finally becomes obscure with increasing $x$, leading to glassy spin dynamics for higher $x$ members. The later observation is also consistent with the assignment of T$_{c2}$ transition to Co$^{3+}$/Mn$^{3+}$ ions since their relative contributions increase (as elucidated from XAS analysis) with increasing $x$. This relative increase leads to emergence of competing Co$^{3+}-$ O$^{2-}-$Co$^{3+}$/Mn$^{3+}-$ O$^{2-}-$Mn$^{3+}$/Mn$^{3+}-$ O$^{2-}-$Mn$^{4+}$ antiferromagnetic superexchange interactions in the cation-disordered phase with increasing $x$. Accordingly, the magnetic moment value measured at $7$ Tesla also decreases steadily with increasing $x$ which is also accompanied by a decrease in the corresponding coercive fields, as seen in Fig.~\ref{fig:S8} of SI \cite{SI}. The increasing magnetic frustrations finally results in glassy spin dynamics above $x = 0.5$, as exemplified by the imaginary part of the AC susceptibility data as shown in \autoref{Fig. (5)}(h) for $x = 0.75$, where clear frequency dependent peaks are observed over a broad temperature window around $75$ K [marked by skewed arrow in \autoref{Fig. (5)}(h)]. The slight increase of T$_{c1}$ for higher $x$ can be explained by considering the increase in average Co-O-Mn bond angle ($\sim 153.6^{\circ}$ in $x = 0$ compared to $\sim 166.8^{\circ}$ in $x = 1$, obtained from RT-NPD) due to reduction in lattice strain. The closer to $180^{\circ}$ the $B–O–B$$^{'}$ angle is, the more symmetric is the crystal structure, and the exchange interactions are stronger \cite{eb8,vasala2015,iba2}. 

With the crossover of the T$_{c2}$ transition of the cation disordered phase into a glassy magnetic transition above $x = 0.5$, the corresponding ordering of the rare-earth moments start becoming apparent at further lower temperatures ($\sim 25$ K), as shown in the DC magnetization data in \autoref{Fig. (5)} and Fig.~\ref{fig:S7} of SI \cite{SI}. The low-temperature ordering of the rare-earth moments is also clearly seen in the real part of the AC susceptibility data for all $x$, as shown in Fig.~\ref{fig:S9} of SI \cite{SI}. At T$_{c1}$ and T$_{c2}$, parallel alignment of the rare-earth moments are driven by the effective magnetic fields due to ferromagnetic alignment of the transition metal cations. Due to such involvement of large rare-earth spin moments at high temperatures, the Curie-Weiss temperatures deviate from the magnetic ordering temperatures of the transition metal cations. At lower temperatures (below $\sim 50$ K), anti-parallel alignment of the rare-earth moments (relative to the TM sublattice) are driven by the $3d-4f$ exchange interactions \cite{Pal2019}. Associated with the glassy dynamics of the transition metal spins corresponding to the B-site cation disordered phase, for $x$ greater than $0.5$, the corresponding rare-earth moments also exhibit glassy spin dynamics around $\sim 25$ K, likely driven by the significant magnetic dilution by Sr ion doping. For $x = 0.75$, both the frequency-dependent dispersions around $\sim 75$ K and $\sim 25$ K exhibit cluster-glass behavior, as evidenced by an unusually large relaxation time ($\tau_0 \approx 10^{-5}$ s), obtained from the power-law fitting (see Fig.~\ref{fig:S10} of SI \cite{SI}) \cite{scale1,scale2} and high Mydosh parameters ($0.105$ and $0.093$ for the $25$ K and $75$ K transitions, respectively) \cite{Mydosh}. Apart from $x = 0.75$, the $25$ K transition is also explicitly visible in $x = 1$ [as seen in \autoref{Fig. (5)}(j)], and a power law fitting of this transition yields similar cluster-glass response ($\tau_0 \approx 10^{-6}$).

\begin{figure}
\scalebox{0.4}
{\includegraphics[width=\textwidth]{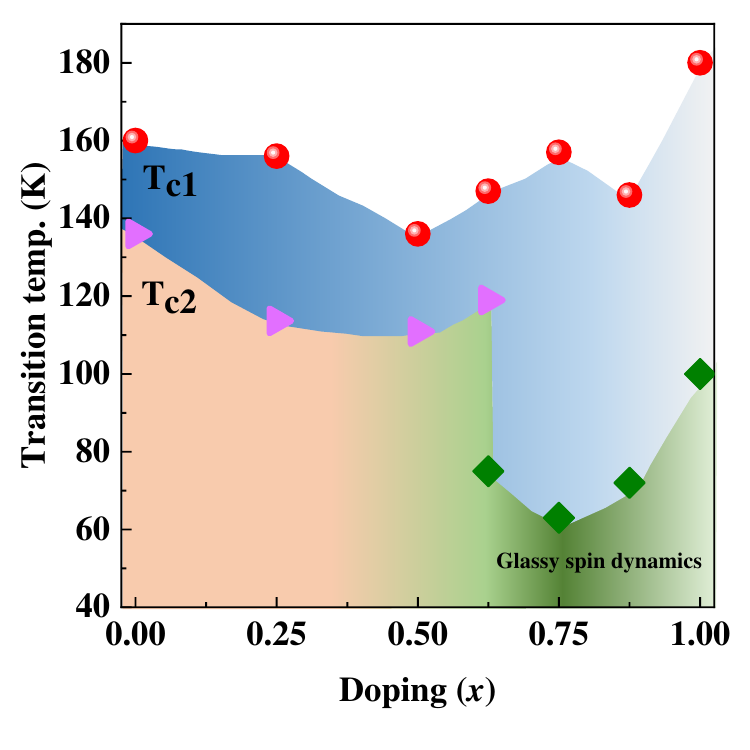}}
\caption{(color online) Magnetic phase-diagram of the Nd$_{2-x}$Sr$_{x}$CoMnO$_{6}$ series showing evolution of transition temperatures as a function of $x$. The color gradient indicates strength of magnetic interactions with deeper color meaning stronger interactions. Blue indicates the ferromagnetic phase after T$_{c1}$ with gradually diminished interaction strength with increasing doping, characterized by the fading blue color. Similarly, the light-orange color indicates the phase after both the ferromagnetic transitions T$_{c1}$ and T$_{c2}$. The green color indicates glassy spin dynamics (strongest around $x = 0.75$) generated by competing ferro and antiferromagnetic interactions.}\label{Fig. (6)}
\end{figure}

\subsection{Exchange bias fields}

\begin{figure*}
{\includegraphics[width=\textwidth]{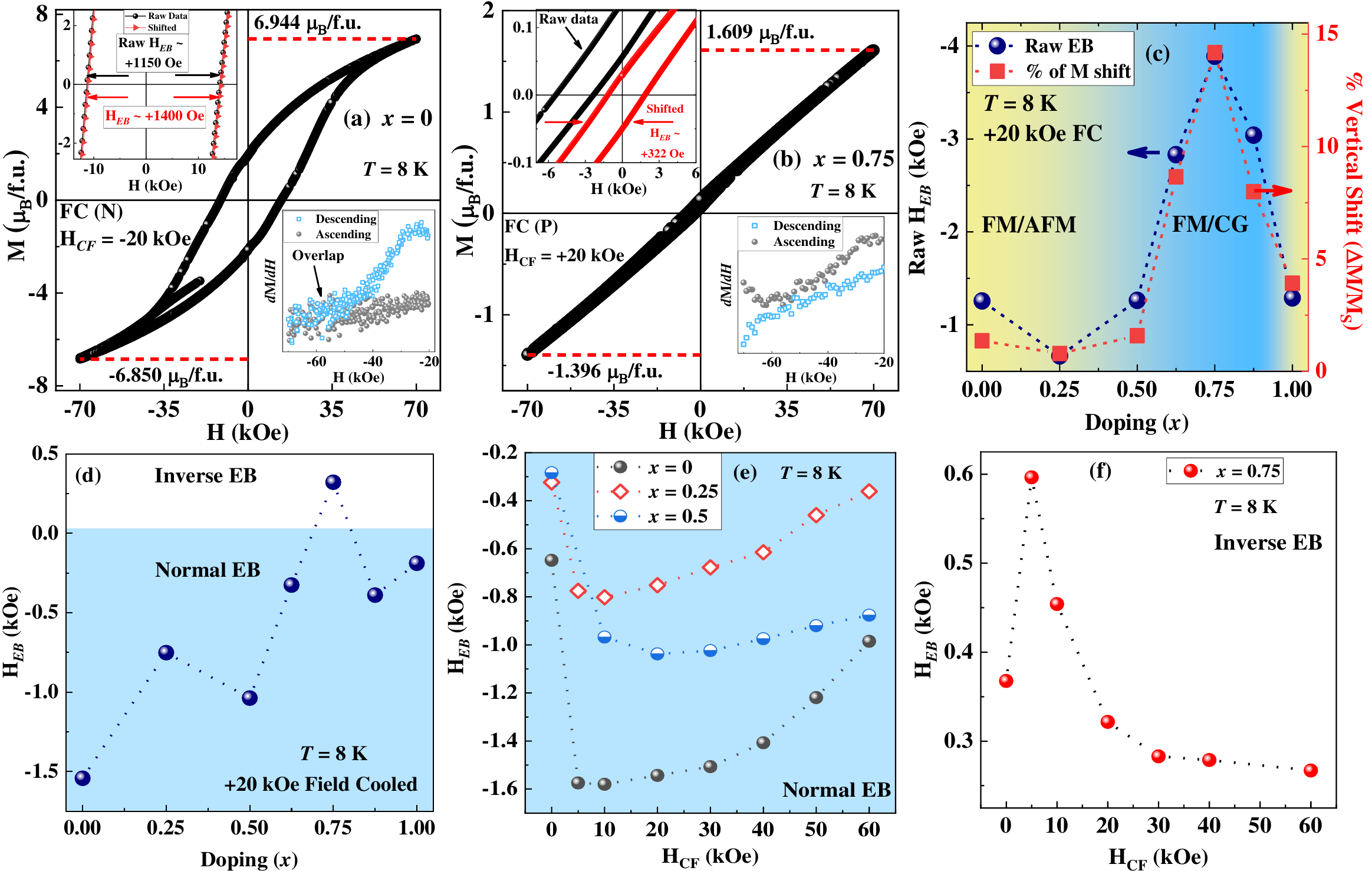}}
\caption{(color online) (a) $8$ K M(H) data of the $x = 0$ compound in FC (N) mode (cooling field $-20$ kOe, CEB). Notice that the loop is slightly vertically shifted. Upper inset shows shifting of the loops along the horizontal axis (raw EB value of $+1150$ Oe) and vertically shifted version of the loop to make the net vertical shift $0$. Notice the change in EB value, now at $+1400$ Oe. Lower inset shows first derivatives of the M(H) curve are starting to align for both ascending and descending branch, indicating the onset of magnetic saturation. (b) $8$ K M(H) data of the $x = 0.75$ compound in FC (P) mode (cooling field $20$ kOe, CEB). Inset shows raw EB value of $-4000$ Oe and vertically shifted version of the loop to make net vertical shift $0$. Notice the sign change in EB value, now only at $+322$ Oe. (c) Variation of both raw CEB data and normalized magnetization shift for the NSCMO series as a function of Sr doping ($x$). Notice how the raw EB values closely follow the $\%$ vertical shift. (d) Background corrected CEB values as a function of Sr doping ($x$). Notice the inverse EB as shown by the $x = 0.75$. (e) Cooling field dependence of EB values for several compositions involving classic FM/AFM type interfaces and AFM interfacial coupling. Note that the ``$0$" cooling field is actually cooling in presence of a trapped magnetic field ($\sim 10$ Oe). (f) Cooling field dependence of EB values for $x = 0.75$ involving FM/glass type interfaces and AFM interfacial coupling. Notice the robust inverse EB shown by $x = 0.75$ at all cooling fields up to $6$ Tesla.}\label{Fig. (7)}
\end{figure*}
As the magnetic phases in Nd$_{2-x}$Sr$_{x}$CoMnO$_{6}$ continue to evolve with $x$ and result in a rich magnetic phase diagram, as seen in \autoref{Fig. (6)}, intriguing EB properties are expected to occur in this series associated with various kinds of interfaces between FM, AFM and glassy magnetic phases. As discussed earlier, to unequivocally establish the EB results, it is imperative to verify whether the sample under investigation has attained magnetic saturation. In this context, singular point determination (SPD) methodology, which involves investigation of the merging of the ascending and descending $dM/dH$ (and higher order derivatives) branches, is usually employed to investigate attainment of magnetic saturation \cite{spd1,spd2,harres2016}. In the event that magnetic saturation is not attained till the highest achievable magnetic field, due to inherently large magnetic anisotropy or disorder, the corresponding M-H loops do exhibit vertical shift, i.e. instead of the centroid of the M-H loop lying on the magnetic field axis [as shown in \autoref{Fig. (2)}(c)], the centroid gets shifted to a finite $\Delta M$ value [as shown in \autoref{Fig. (2)}(d) or the vertically shifted loop of \autoref{Fig. (2)}(e)]. The vertical shift magnitude normalized with the high field magnetization ($M_S$) then acts as an indicator of the strength of magnetic pinning/disorder. In our samples, the intrinsic EB fields (henceforth referred to as ``H$_{EB}$") have been obtained by correcting for such vertical shifts in the M-H loops using the methodologies previously established by us \cite{mishra}. In the following, we refer to the EB field as ``Raw H$_{EB}$'' which are directly read from vertically shifted M-H loops (i.e. raw data) without correcting for the vertical shifts. \par
We first discuss whether the Nd$_{2-x}$Sr$_{x}$CoMnO$_{6}$ samples do exhibit any spontaneous exchange bias (SEB). In compounds exhibiting the novel SEB effect, it is possible to introduce the necessary unidirectional magnetic anisotropy at the interface between the FM and AFM regions during the  P or  N field ramping procedure in the M-H measurement (as detailed in the methodology section) even in absence of any finite magnetic field during cooling of the sample, where the cooling field is usually necessary to introduce such an anisotropy in case of systems exhibiting conventional exchange bias (CEB). We first discuss the results for the $x = 0$ compound. Interestingly, even after ZFC, finite H$_{EB}$ field (raw H$_{EB}$ $\sim 500$ Oe; H$_{EB}$ $\sim 700$ Oe) seems to be realized and the sign of the apparent SEB also changes between P and N modes of M-H measurement [Fig.~\ref{fig:S12}(a) of SI \cite{SI}]. These results do seem to suggest that $x = 0$ sample exhibits SEB effect. However, a closer inspection elucidates that the corresponding virgin magnetization curves do not start from zero magnetization, suggesting the presence of small trapped fields within the superconducting magnet coil during the ZFC procedure. To investigate the presence of SEB, it is important to have a precise control on the sign of the trapped magnetic field (which can be verified through the sign of the dc magnetic susceptibility in the temperature window where the samples are paramagnetic) and keep the trapped magnetic field as small ($\sim 10$ Oe) and close as possible (for all the samples) following the established protocols \cite{mishra}. We note that the EB arising solely due to cooling in presence of a trapped magnetic field is hereby referred to as CEB instead of SEB. To explore whether the $x = 0$ sample exhibits SEB along with CEB, we performed control experiments by fixing the sign of the trapped magnetic field and then proceed with the M-H measurements using both the P and N field sweep modes. For example, if the sign of the trapped magnetic field is positive, the H$_{EB}$ in the P sweep mode of M-H measurement will be (CEB$+$SEB) and the H$_{EB}$ in the N sweep mode of M-H measurement will be (CEB$-$SEB), assuming the magnitude of the trapped magnetic field do not vary much between these two runs (which was found in \cite{mishra}). The SEB, if present, can be obtained as a difference between the H$_{EB}$ values of the above P and N M-H measurements, as illustrated in Fig.~\ref{fig:S11} of SI \cite{SI}. The value of SEB for $x = 0$, estimated using the above methodology, is found to be negligible ($\sim 100$ Oe, which is comparable to the associated error bar since the field step size is $50$ Oe for our measurements). Similarly, the SEB is found to be absent (or negligibly small) by employing the above methodologies for higher $x$ members also. The apparent raw SEB, although found to be large for some intermediate compositions (raw H$_{EB}$ $\sim 1400$ Oe; H$_{EB}$ $\sim 400$ Oe for $x = 0.75$) [Fig.~\ref{fig:S12}(b) of SI \cite{SI}], is understood to arise due to cooling in presence of finite trapped magnetic fields and, thus, be referred to as CEB in the following discussions. \par
Intriguingly, the $20$ kOe or $2$ Tesla field-cooled (FC) ``Raw H$_{EB}$'' exhibit a clear non-monotonic dependency on $x$ in Nd$_{2-x}$Sr$_{x}$CoMnO$_{6}$, with a clear maximum ($\sim 4000$ Oe) around $x = 0.75$, as seen in the left axis in \autoref{Fig. (7)}(c). Interestingly, similar non-monotonic maximization of ``Raw H$_{EB}$'' at some intermediate $x$ composition have also been observed in few other hole-doped double perovskite compounds \cite{cmo2,asd_x1,sahoo2021}, as discussed earlier. Such phenomena have been explained in terms of the optimization of FM/AFM interface with gradual hole doping in the system. This conventional interpretation states that as hole doping increases, the antiferromagnetic (AFM) interaction becomes more pronounced inside the ferromagnetic (FM) matrix due to antisite disorder. Consequently, this leads to the formation of a greater number of interfaces exhibiting AFM/FM-type (or SG/FM, AFM/SG, etc.) characteristics. However, sometimes, once doping reaches a certain threshold, the prominence of FM interactions diminishes, resulting in a gradual loss of interfaces. Therefore, at a certain level of doping, there exists an optimal balance between the quantities of ferromagnetic (FM) and antiferromagnetic (AFM) components, resulting in the maximum generation of exchange bias (EB). These results/interpretations are often corroborated with the classical Meiklejohn and Bean relation \cite{ebg2,mb1}, which was originally proposed for the FM/AFM bilayer films. However, as explained earlier, these ``Raw H$_{EB}$'' values are generally accompanied by large vertical shifts in the hysteresis loops and do not represent intrinsic EB fields at some magnetic interfaces. \par
The M-H curve of $x = 0$ exhibits negligible vertical shift, whereas the M-H curve of $x = 0.75$ exhibits unusually large vertical shift, as seen in \autoref{Fig. (7)}(a) and \autoref{Fig. (7)}(b), respectively. SPD methodology also suggests that $x = 0$ is close to magnetic saturation, as opposed to $x = 0.75$ (seen in the corresponding insets). Strikingly, the associated normalized vertical shifts of the centriods of the corresponding M-H loops ($\Delta M/M_s$) do exhibit an identical non-monotonic dependency on $x$ as the ``Raw H$_{EB}$'' values, as shown on the right-axis in \autoref{Fig. (7)}(c). This comparison between $x$ dependence of ``Raw H$_{EB}$'' and the vertical shifts $\Delta M/M_s$ clearly elucidate that the ``Raw H$_{EB}$'' values are not intrinsic to any magnetic interfaces and arise from minor-loop related effects. The intrinsic H$_{EB}$ obtained after correcting for the vertical shifts of the corresponding M-H curves are plotted for a specific cooling field ($+20$ kOe) as a function of $x$ in \autoref{Fig. (7)}(d). Unlike in \autoref{Fig. (7)}(c), which suggests that ``Raw H$_{EB}$'' field increases with $x$ till some intermediate composition $x = 0.75$ followed by a decrease thereafter, the H$_{EB}$ continues to decrease with increasing $x$ or increasing anti-site disorder. Strikingly, the $x = 0.75$ composition, which exhibits significant magnetic glassiness, is found to exhibit inverse H$_{EB}$ (for example, H$_{EB} \sim +322$ Oe for a cooling field of $+20$ kOe). Importantly, the inverse H$_{EB}$ (IEB) is robust for cooling fields till $60$ kOe or $6$ Tesla [shown in \autoref{Fig. (7)}(f)]. \par
To delve into the nature of magnetic interfaces in these compositions and also to determine the sign of exchange coupling across these interfaces, we have shown the minor loop corrected EB values as a function of cooling fields (H$_{CF}$) in \autoref{Fig. (7)}(e) and \autoref{Fig. (7)}(f) as well as in Fig.~\ref{fig:S13} of SI \cite{SI}. It can be seen that for $x = 0$, $x = 0.25$, and $x = 0.5$ the H$_{EB}$ increases at first with applied H$_{CF}$ but later decreases with increasing H$_{CF}$. Usually, at small cooling field, the field can increase the number of spins at the interface aligned along the external magnetic field and reduce the effect of averaging of the anisotropy due to randomness, which enhances the spin coupling between the AFM/FM interface and hence the value of H$_{EB}$. As the cooling field further increases, the number of spins is saturated, which results in the saturation of H$_{EB}$. This behavior of H$_{EB}$ can be mathematically explained in terms of a well-known model proposed by D. Niebieskikwiat and M. B. Salamon \cite{quiet} for phase-separated systems consisting of single-domain FM clusters embedded in the AFM matrix:
\begin{equation}
\text{H}_{EB} \propto J\{[J\mu_0/(g\mu_\text{B})^2] \, L(\mu H_{\text{cool}}/k_B T_f) + H_{\text{cool}}\},
\end{equation}
where \textit{J} is the interface exchange constant, \textit{g} $= 2$ is the gyromagnetic factor, $\mu_\text{B}$ is the Bohr magneton, \textit{L} denotes the Langevin function, $\mu = N \mu_0$ is the magnetic moment of the FM clusters with N number of spins, and $T_f$ is the spin freezing temperature. Equation ($1$) has been frequently used for evaluation of the FM cluster size in a variety of EB phase-separated systems, such as manganites and cobaltites. In this equation, the competition between the exchange interaction and the cooling field becomes evident. For small H$_{\text{cool}}$ the first term usually dominates, and H$_{E}$ depends on $J_i^2$. However, for large cooling fields the second term ($\propto J_i$) becomes important, and for $J_i < 0$ the absolute value of H$_{EB}$ could decrease or even more, H$_{EB}$ could change sign, as observed in previous works  \cite{salamon1,eb9}. In the context of current results, the decrease in H$_{EB}$ indicates antiferromagnetic type interfacial coupling $(J_i < 0$) between the FM and AFM layers.\par
The $x = 0.75$ composition, which shows the highest non-corrected EB (``Raw H$_{EB}$") of $\sim 4$ kOe at a cooling field of $6$ Tesla and a sweeping field of $7$ Tesla, exhibits robust IEB after the minor loop correction. This composition never switches to NEB, even with application of high cooling field of $6$ Tesla. Similar type of behavior has been shown by Gd$_{2}$CoRuO$_{6}$, LuFe$_{0.5}$Cr$_{0.5}$O$_{3}$ and Gd$_2$Co$_{0.5}$Mn$_{1.5}$O$_{6}$ etc. \cite{eb3,eb6,peb1,peb5}. This ``exponential" decrease is typical of FM/Glass type interfaces \cite{sgieb6,fmsg1}. The reason for such type of behavior (``IEB preservation") is not well understood till date, although it has been attributed to various scenarios. For example, according to J. Nogu\'es \textit{et al.} \cite{eb9} rough interfaces can generate spatially varying AFM and FM couplings. Such mixed couplings can produce robust inverse EB. On the other hand, according to Canglong Li \textit{et al.} \cite{peb1}, the reduction in the number of AFM spins provided for pinning is responsible for the decrease in amplitude of H$_{EB}$ with increasing H$_{CF}$. In the present context, it seems that strong glassy phase present (confirmed by the AC susceptibility data as well as an exceedingly high vertical magnetization shift of $\sim 14$\%) in this composition makes FM/glass-type interface prevalent at all cooling fields, as such high vertical magnetization shift can potentially originate from the existence of glassy phase at the interface \cite{vms1,vms2}. In the low cooling field, the sharp increase in EB field is due to the gradual saturation of FM layer, while in high cooling field, the subsequent decrease is due to the polarization of the glassy phase facilitated by cooling field, resulting in potential loss of EB interfaces, which could explain the diminish in IEB value. Finally, significant magnetic disorder and weakening of the FM order in $x = 1$ results in it's small H$_{EB}$ fields (as seen in Fig.~\ref{fig:S13} of SI \cite{SI}), potentially originating from FM/AFM type interfaces.

\section{Conclusions}
At the outset, the main aim of this study was to investigate the dependence of intrinsic exchange bias fields on the extent of anti-site disorder in a double perovskite series of compounds. Sr-doped Nd$_{2}$CoMnO$_{6}$ constitutes a unique family, where with increasing anti-site disorder, a gradual transition is observed from a B-site cation ordered $P2_1/n$ structure to B-site cation disordered $R\overline{3}c$ structure. Mainly driven by localization of the doped holes on the Co-sites, a series of magnetic phases consisting of various weightages of FM, AFM and glassy magnetic phases depending on $x$ is obtained which leads to an exotic magnetic phase diagram. Notably, strong pinning centers and presence of disordered magnetic regions, which are more prevalent for intermediate $x$ compositions, lead to strong magnetic anisotropy that lead to large minor-loop effects in the associated low-temperature M-H loops. Such minor loop effects can be erroneously interpreted as exchange-bias fields, which then exhibits non-monotonic dependency (going through a well-defined maxima for some intermediate values of hole doping). Further, cooling of samples in presence of finite trapped magnetic fields can lead to exchange bias fields, which can be erroneously interpreted as spontaneous exchange bias. Investigation of the intrinsic exchange-bias fields which were obtained after correcting for the vertical shifts in the associated M-H loops demonstrate that such a non-monotonic dependency of H$_{EB}$ on the extent of anti-site disorder, where exchange bias fields increase with increasing the extent of anti-site disorder is not an intrinsic property of the double perovskite series. Interestingly, driven by the rich magnetic phase diagram of this series, a transition from normal to inverse exchange bias fields is observed around compositions where a predominance of glassy magnetic phases along with long-range magnetic order is concomitantly observed, though spatially phase separated. Hole doped double perovskite series of compounds thus, seem to be an ideal playground to look for rich magnetic and exchange bias properties.  

\section{Acknowledgements}
K.P.I. acknowledges financial support from the Ministry of Education (MoE), Government of India. The authors acknowledge the experimental facilities at the Central Research Facility (CRF), and the Department of Mechanical Engineering, IIT Kharagpur, which were utilized for various measurements. We also acknowledge DESY (Hamburg, Germany), a member of the Helmholtz Association HGF, for providing the experimental facilities where parts of this research were carried out at the P04 beamline of PETRA III using the Max-P04 instrument under Proposal ID I-20250586. Financial support from the Department of Science \& Technology (Government of India) through the India@DESY collaboration is gratefully acknowledged. The authors further thank Prof. D. D. Sarma for valuable discussions and insightful suggestions.

\onecolumngrid

\phantomsection

\section*{Supplemental Figures and Captions}

\setcounter{figure}{0}
\renewcommand{\thefigure}{S\arabic{figure}}
\renewcommand{\theHfigure}{S\arabic{figure}}

\begin{figure}[h!]
\centering
\scalebox{0.8}
{\includegraphics[width=\textwidth]{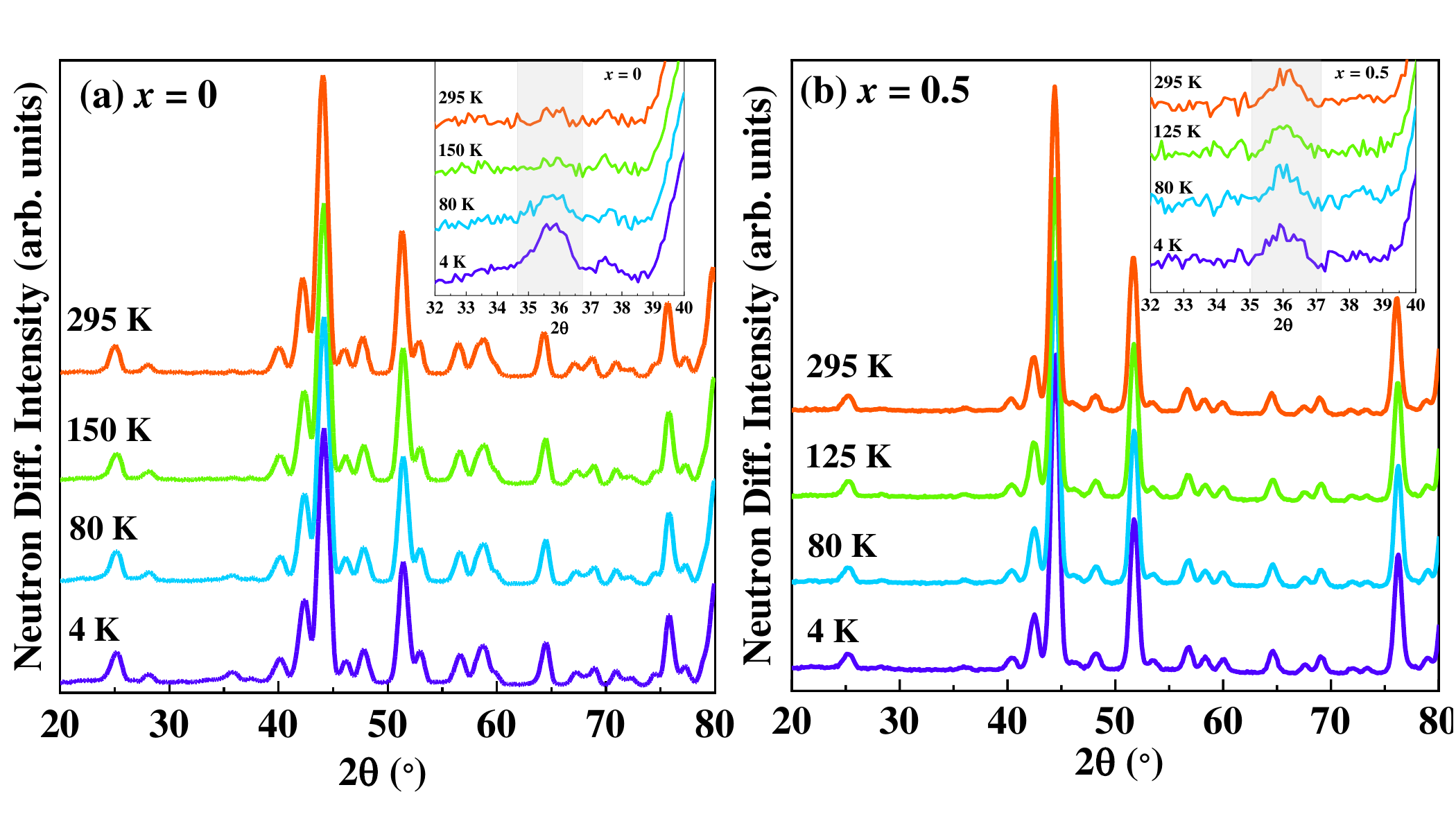}}
\caption{(Color online) The temperature-dependent NPD data show no evidence of a structural phase transition down to 4~K for any of the investigated $x$ compositions. Representative diffraction patterns for (a) $x = 0$ and (b) $x = 0.5$ are shown. The inset in (a) illustrates the evolution of the most intense magnetic Bragg peak for $x = 0$. In contrast, the inset in (b) shows that the corresponding magnetic peak becomes obscured by the emergence of nuclear reflections at nearly the same diffraction position for $x = 0.5$, making it difficult to reliably determine the magnetic structure and the associated ordered magnetic moments.}
\label{fig:S1}
\end{figure}

\begin{figure}[h!]
\centering
\scalebox{0.8}
{\includegraphics[width=\textwidth]{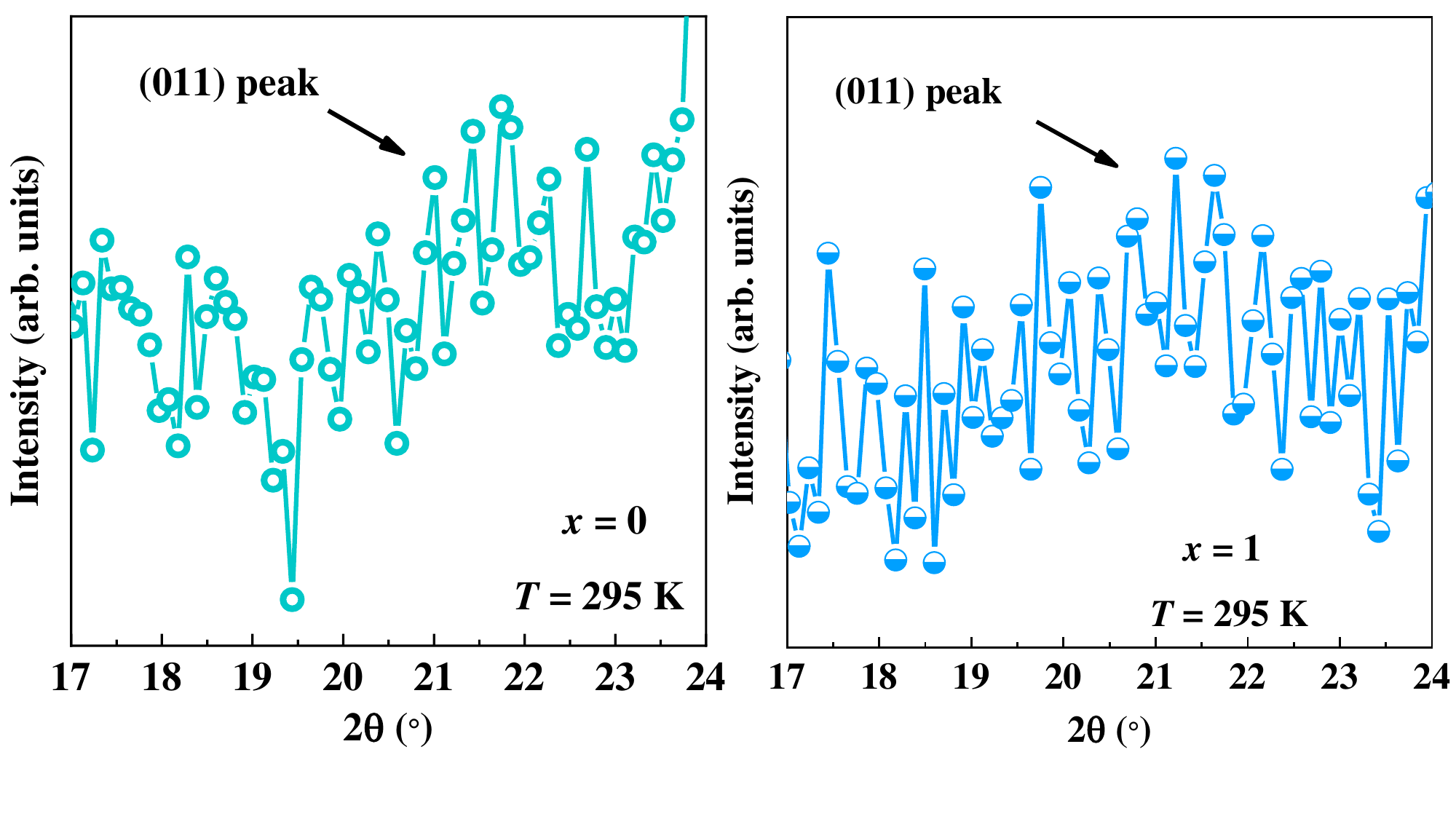}}
\caption{(Color online) Room-temperature (RT) neutron diffraction data showing the superlattice reflections for $x = 0$ and $x = 1$. The superlattice peak for $x = 0$ is slightly more intense than that for $x = 1$, suggesting a reduction in long-range cation ordering and an increased degree of disorder in the $x = 1$ composition.}
\label{fig:S2}
\end{figure}

\begin{figure}[h!]
\centering
\scalebox{0.5}
{\includegraphics[width=\textwidth]{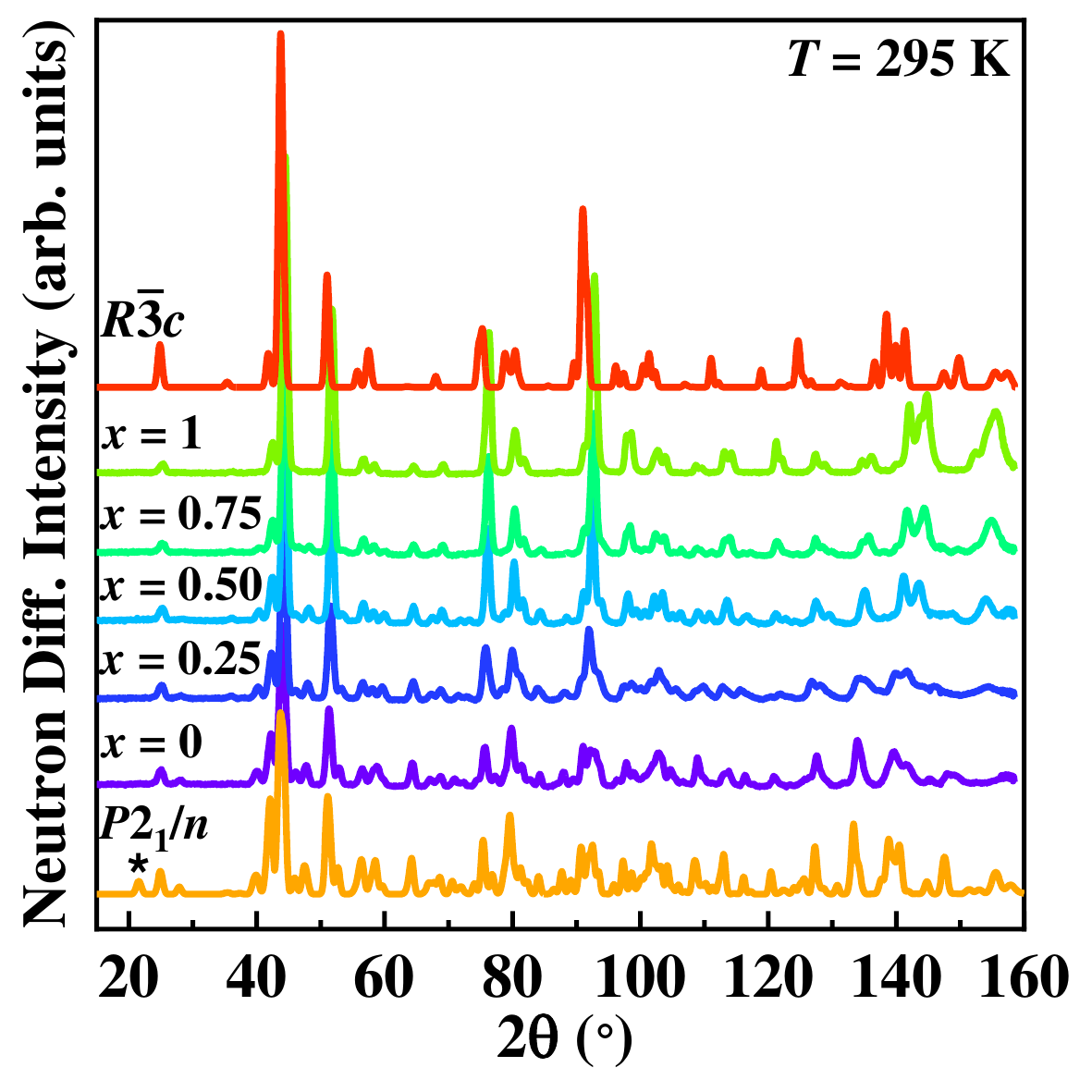}}
\caption{(Color online) Room-temperature (RT) neutron powder diffraction (NPD) patterns of the NSCMO series. The asterisk (*) marks the (011) superlattice reflection. The reference peak positions were obtained from the ICSD database.}
\label{fig:S3}
\end{figure}

\begin{figure}[h!]
\centering
\scalebox{0.5}
{\includegraphics[width=\textwidth]{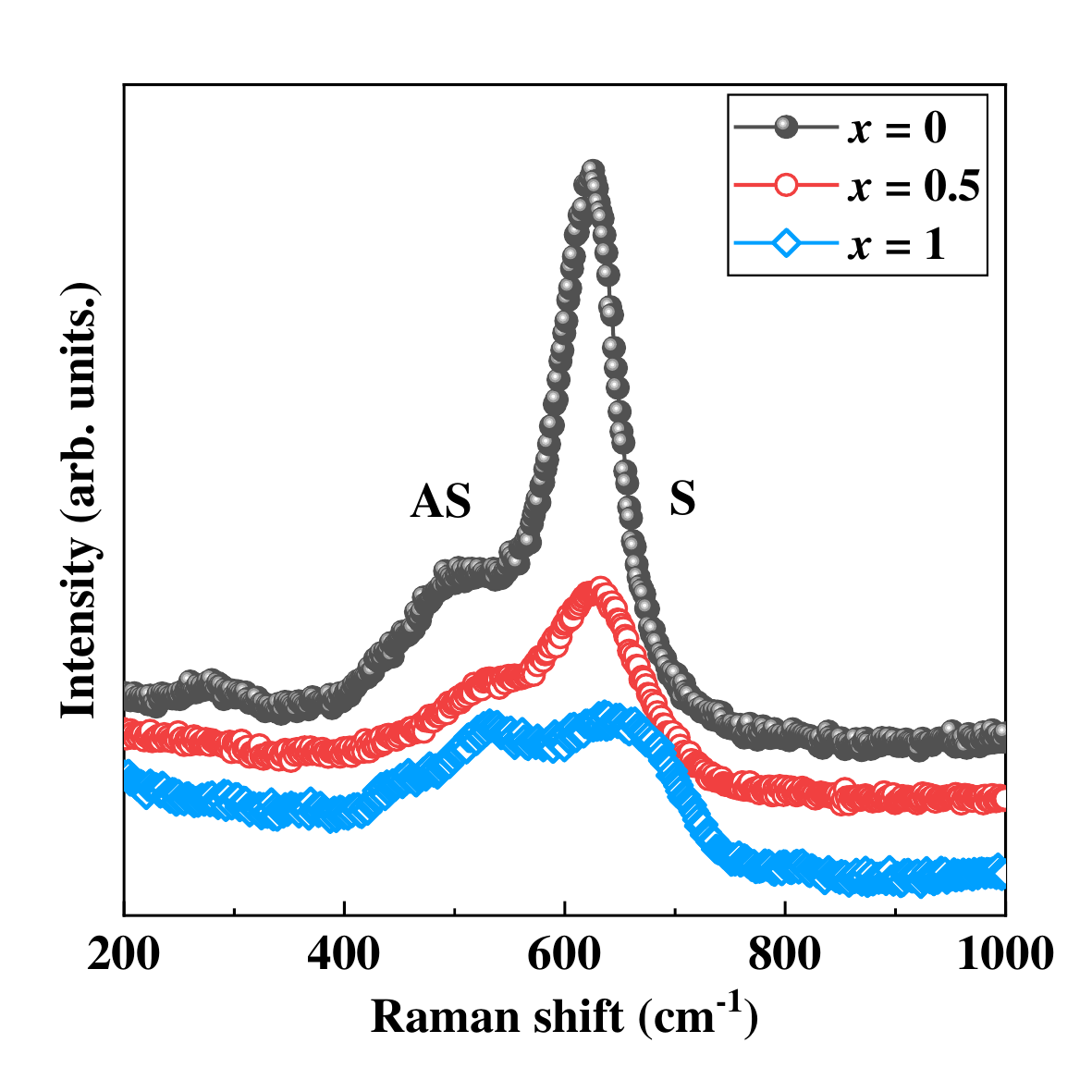}}
\caption{(Color online) Stretching (S) and anti-stretching (AS) Raman vibrational modes for $x = 0$, $0.5$, and $1$. A systematic reduction in peak intensity and an increase in Raman mode broadening are observed with increasing $x$ (hole doping), indicating enhanced lattice disorder.}
\label{fig:S4}
\end{figure}

\begin{figure}[h!]
\centering
\scalebox{0.8}
{\includegraphics[width=\textwidth]{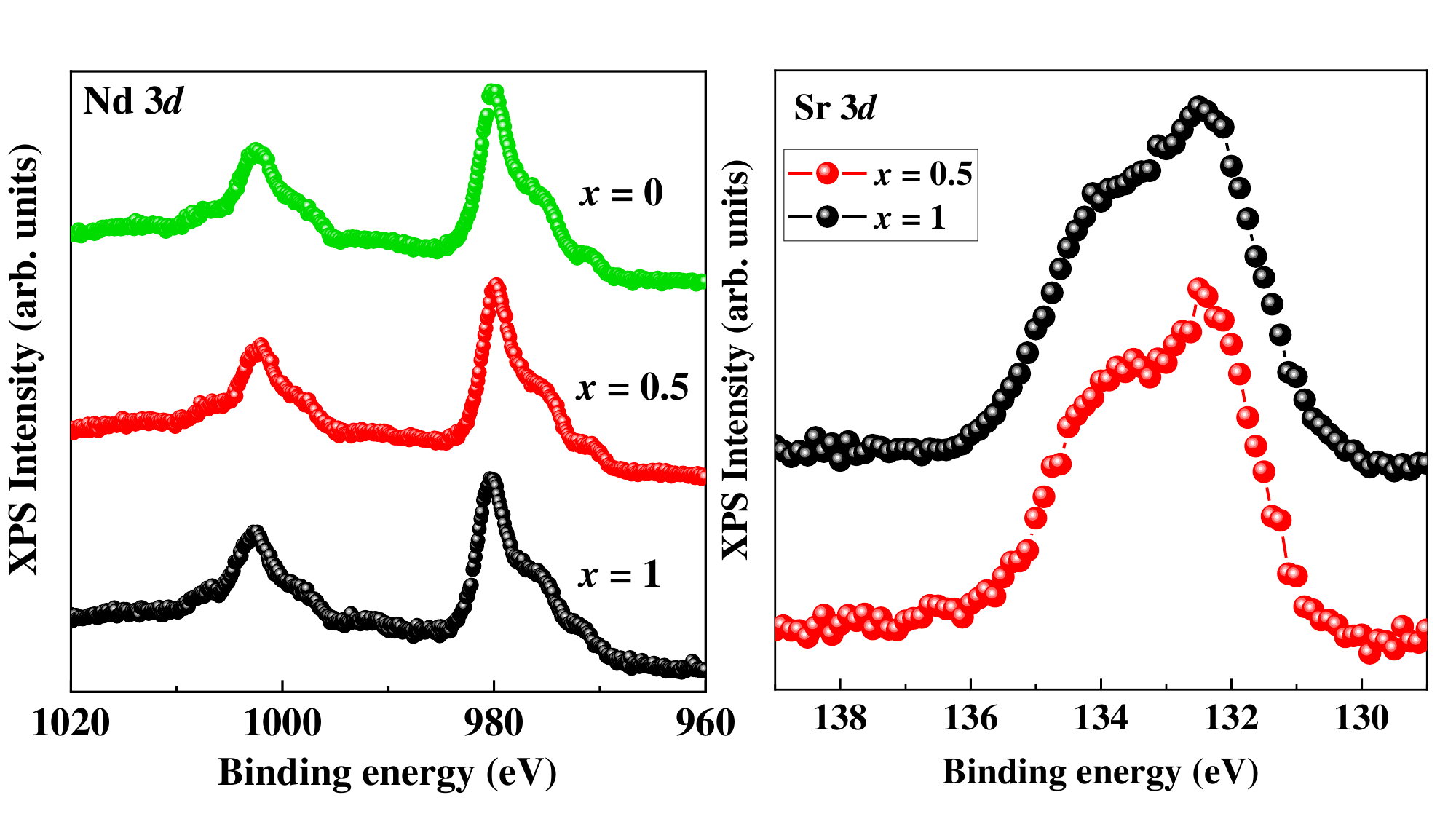}}
\caption{(Color online) The 3$d$ XPS spectra of Nd and Sr indicate stable Nd$^{3+}$ and Sr$^{2+}$ oxidation states, respectively, for all compositions ($x = 0$ [Nd-only], $0.5$, and $1$).}
\label{fig:S5}
\end{figure}

\begin{figure}[h!]
\centering
\scalebox{0.8}
{\includegraphics[width=\textwidth]{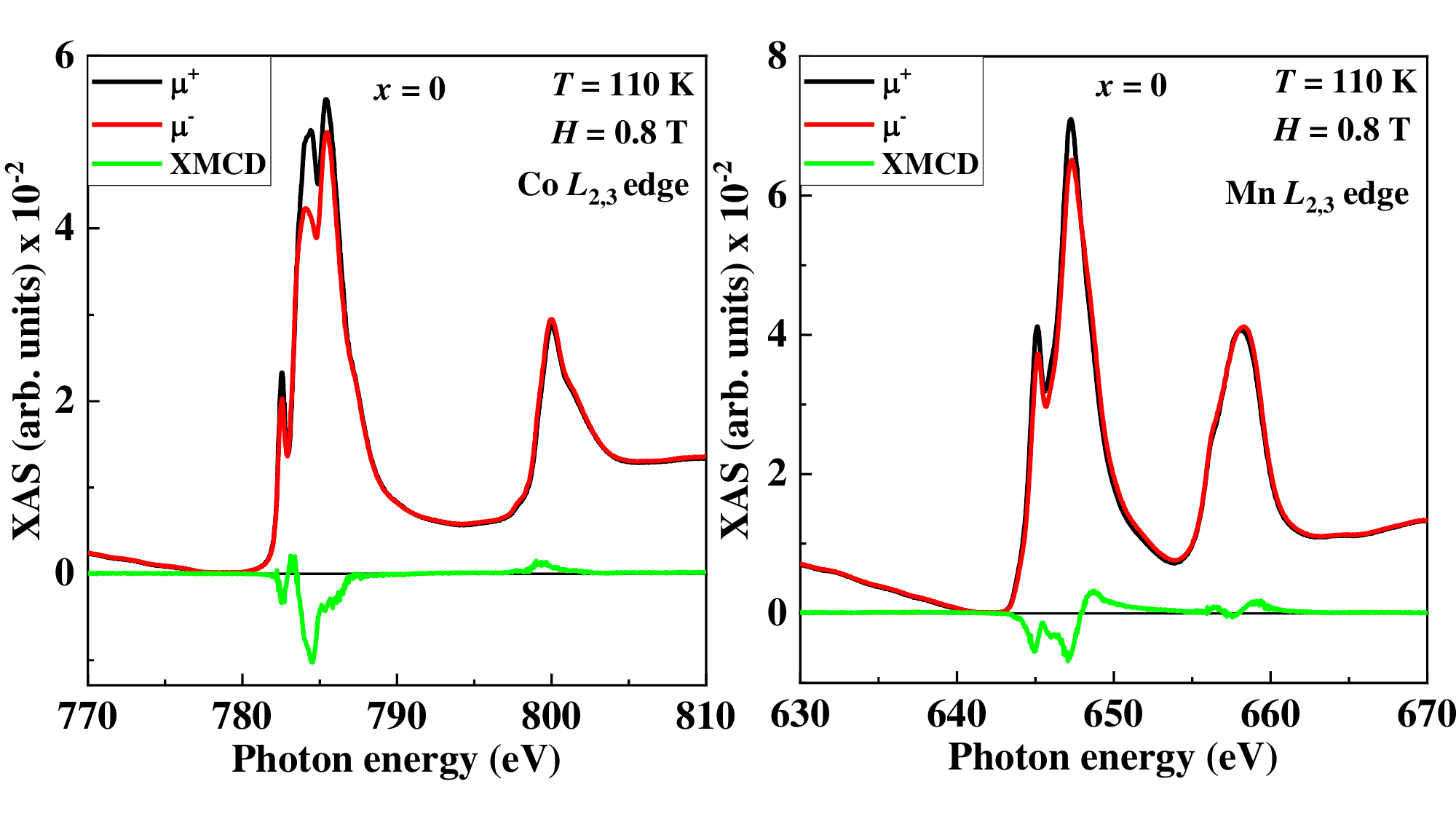}}
\caption{(Color online) XMCD spectra of Co and Mn for $x = 0$ showing the same sign at the $L_3$ absorption edge, indicating parallel alignment of the Co and Mn magnetic moments and, consequently, a ferromagnetic exchange interaction between them.}
\label{fig:S6}
\end{figure}

\begin{figure}[h!]
\centering
\scalebox{0.8}
{\includegraphics[width=\textwidth]{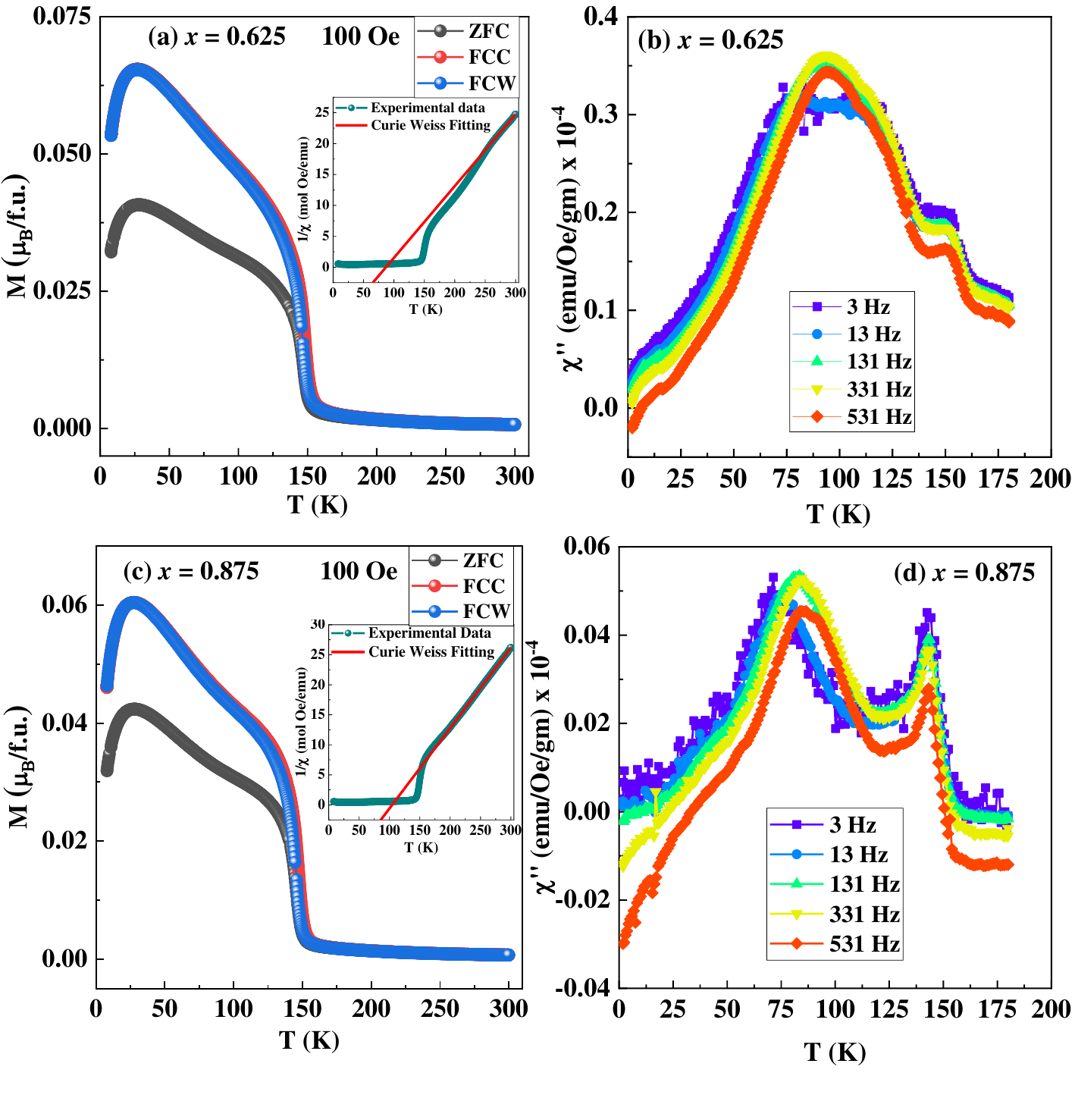}}
\caption{(Color online) Temperature-dependent magnetization for $x = 0.625$ and $0.875$.}
\label{fig:S7}
\end{figure}

\begin{figure}[h!]
\centering
\scalebox{0.5}
{\includegraphics[width=\textwidth]{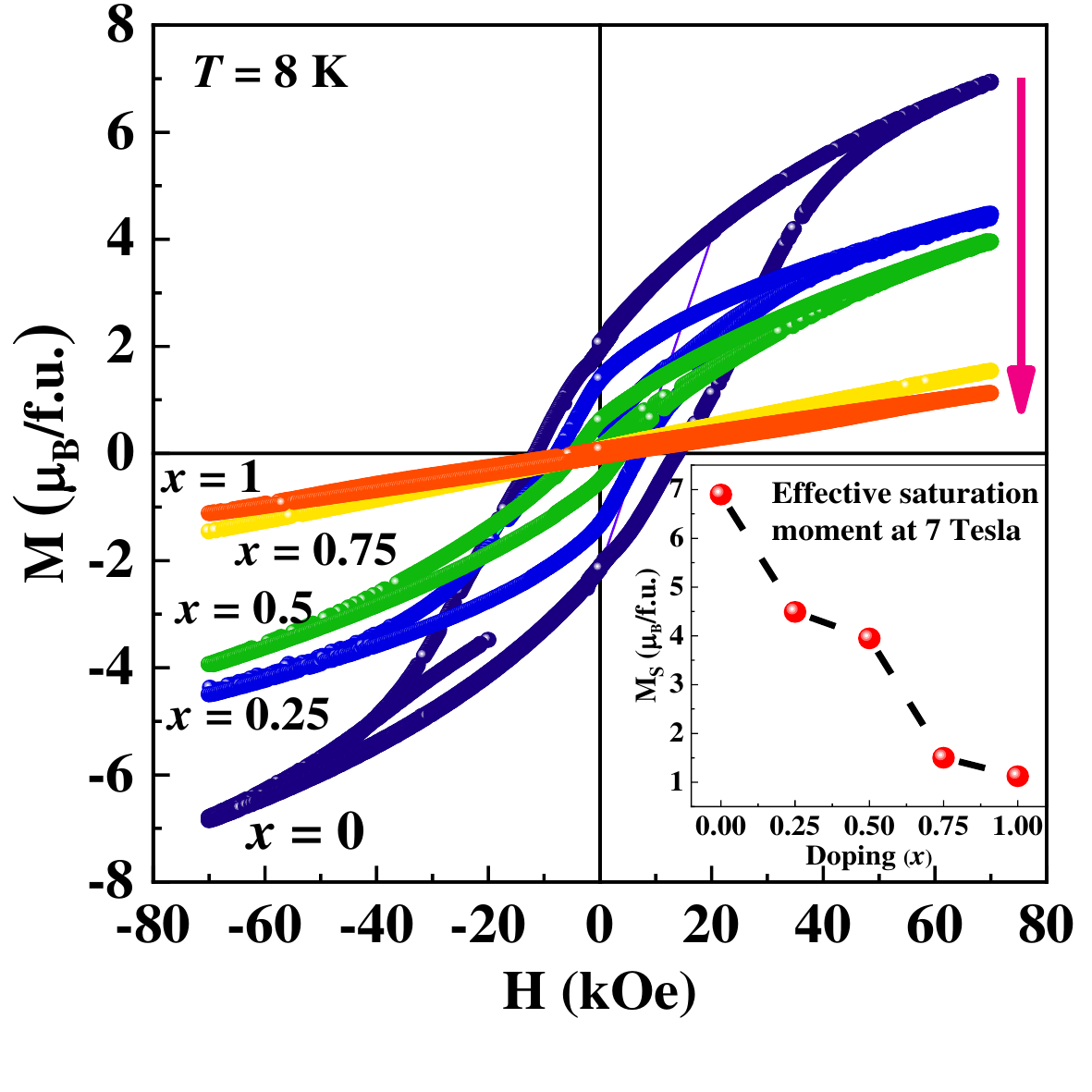}}
\caption{(Color online) Progressive reduction in the saturation magnetization and coercive field with increasing hole doping, consistent with the enhanced degree of disorder.}
\label{fig:S8}
\end{figure}

\begin{figure}[h!]
\centering
\scalebox{0.6}
{\includegraphics[width=\textwidth]{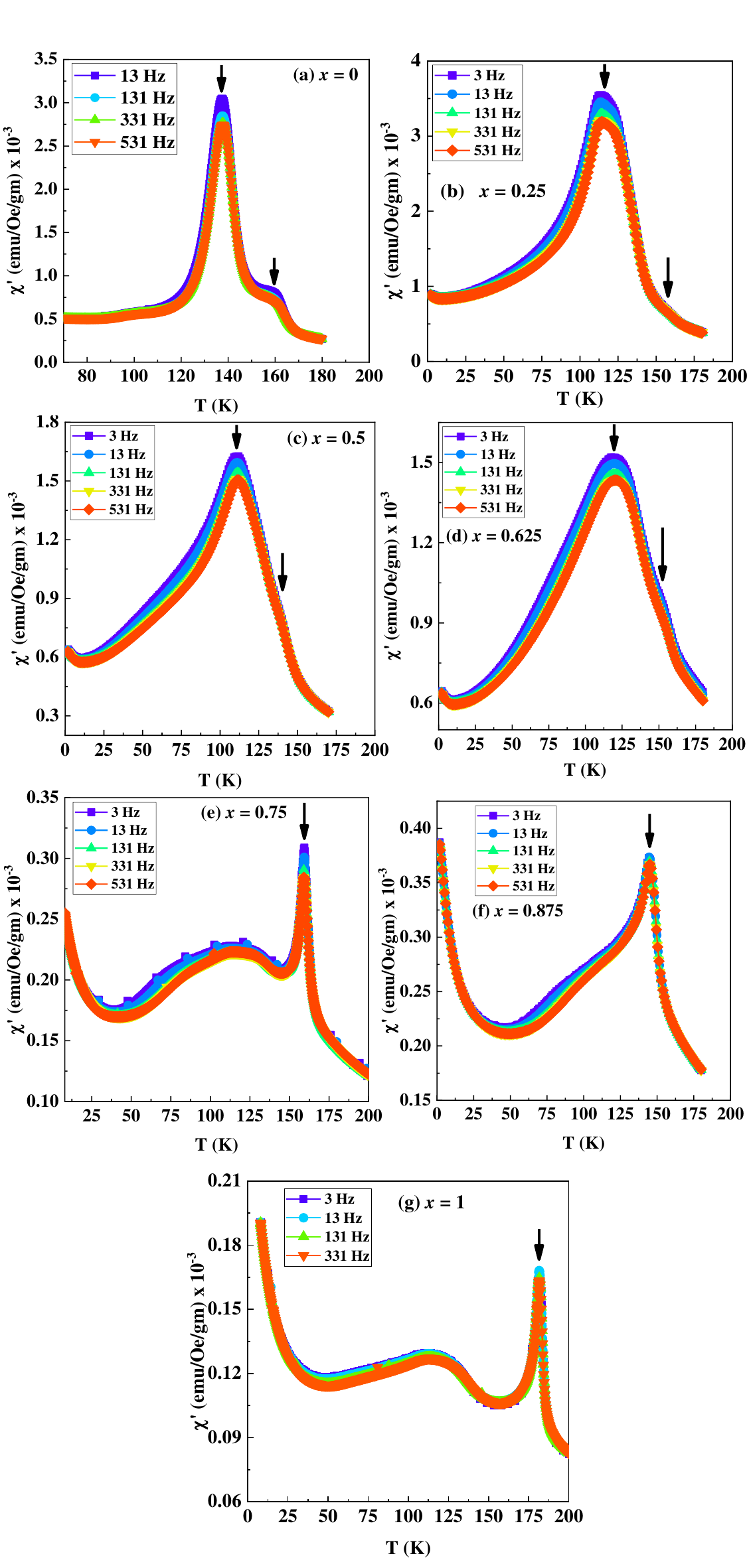}}
\caption{(Color online) AC magnetic susceptibility (real component, $\chi'$) of the NSCMO series.}
\label{fig:S9}
\end{figure}

\begin{figure}[h!]
\centering
{\includegraphics[width=\textwidth]{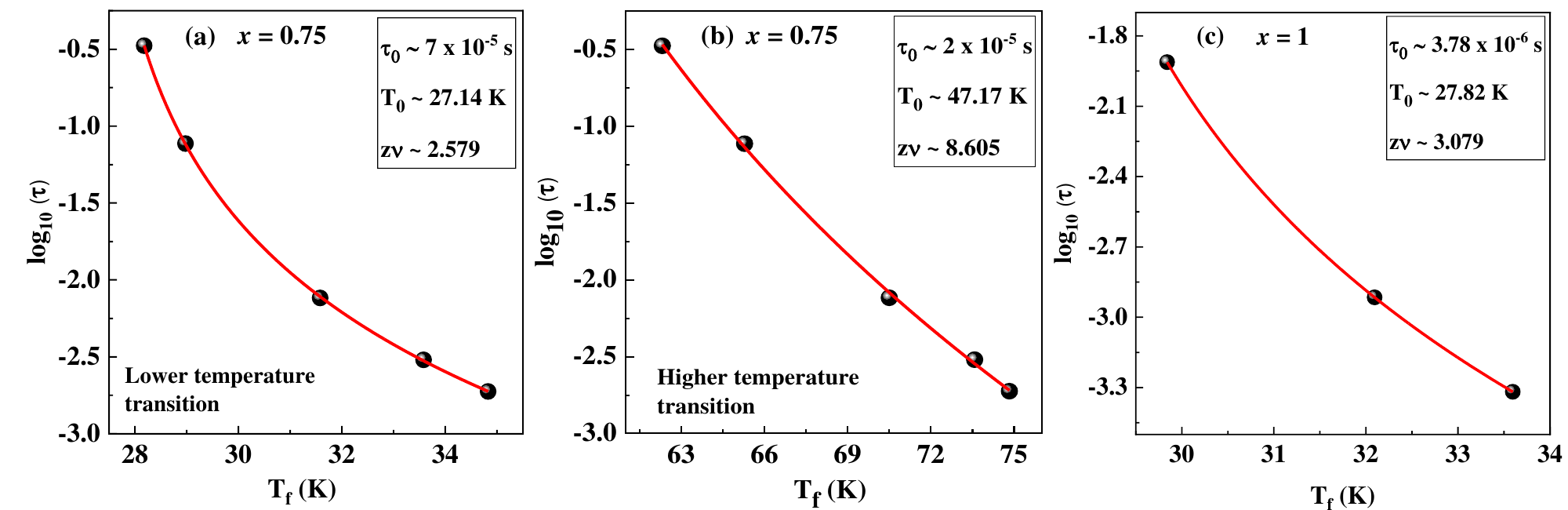}}
\caption{(Color online) Power-law fitting of the AC susceptibility data for (a) and (b) $x = 0.75$ and (c) $x = 1$.}
\label{fig:S10}
\end{figure}

\begin{figure}[h!]
\centering
\scalebox{0.7}
{\includegraphics[width=\textwidth]{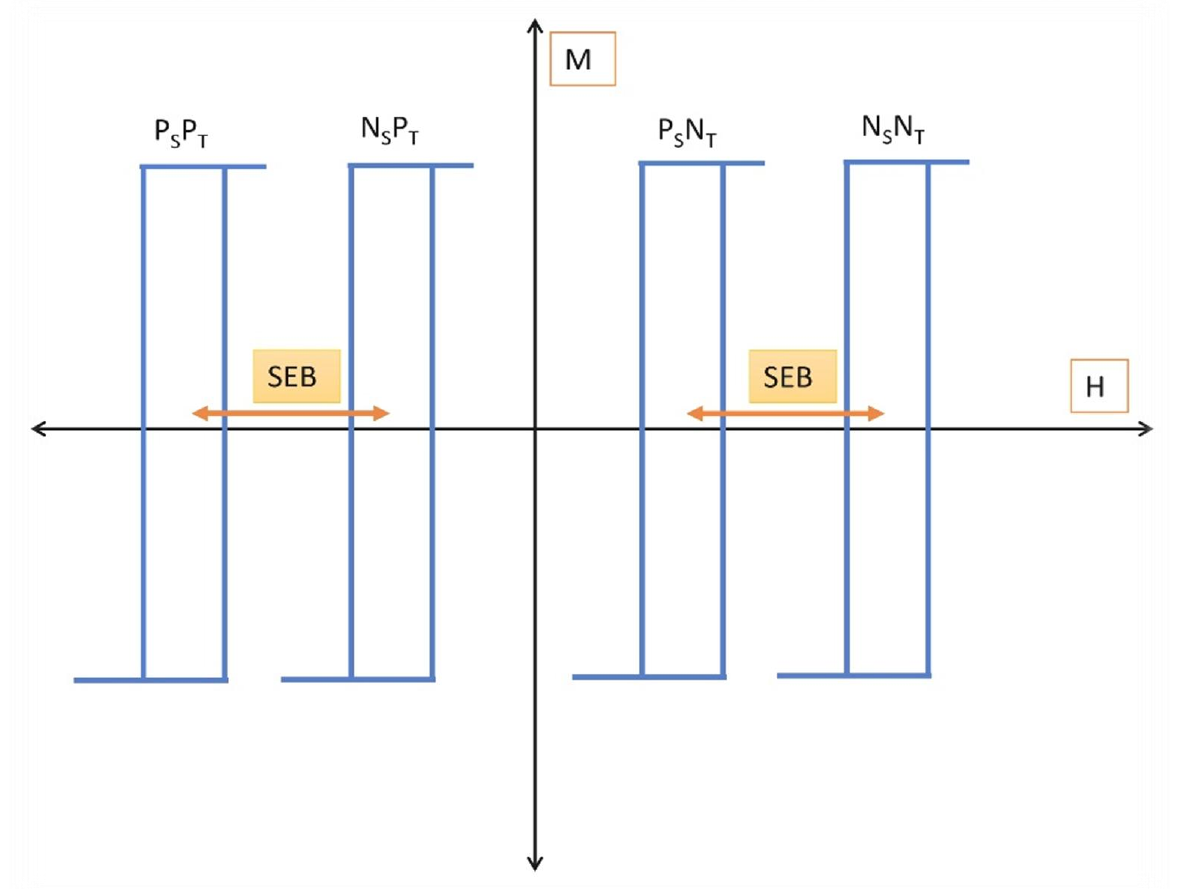}}
\caption{(Color online) Schematic illustrating the methodology used to determine the true spontaneous exchange bias (SEB) in the presence of trapped magnetic fields. Here, $(P/N)_S$ represents a positive/negative field sweep direction (P/N mode), whereas $(P/N)_T$ represents a positive/negative trapped field. The procedure assumes conventional exchange-bias behavior, such that the hysteresis loop shifts opposite to the direction of the applied cooling field.}
\label{fig:S11}
\end{figure}

\begin{figure}[h!]
\centering
\scalebox{0.8}
{\includegraphics[width=\textwidth]{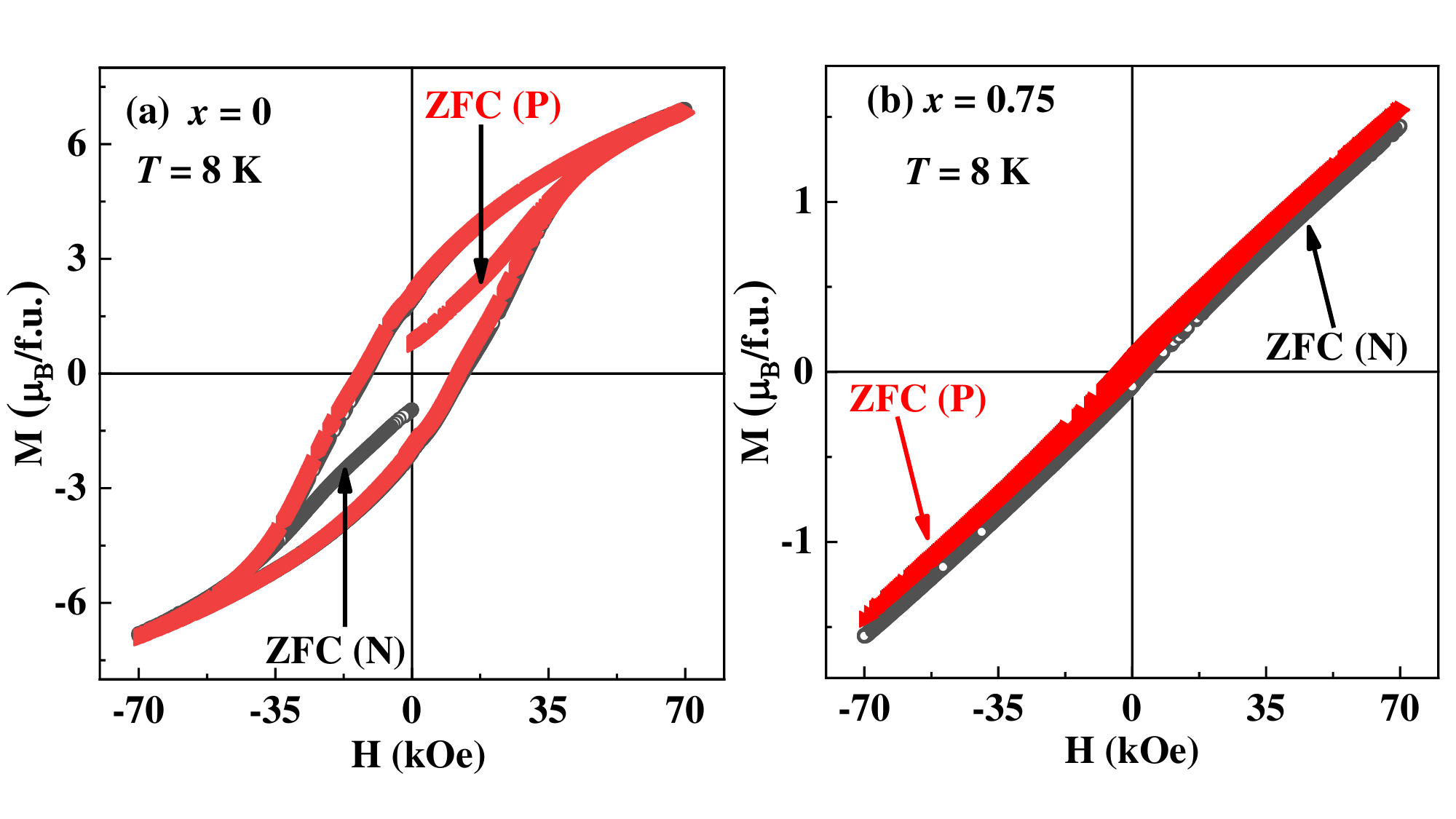}}
\caption{(Color online) “ZFC” $M$--$H$ curves for (a) $x = 0$ and (b) $x = 0.75$, showing different starting points of the virgin magnetization curve, which are indicative of the polarity of the trapped magnetic field.}
\label{fig:S12}
\end{figure}

\begin{figure}[h!]
\centering
\scalebox{0.5}
{\includegraphics[width=\textwidth]{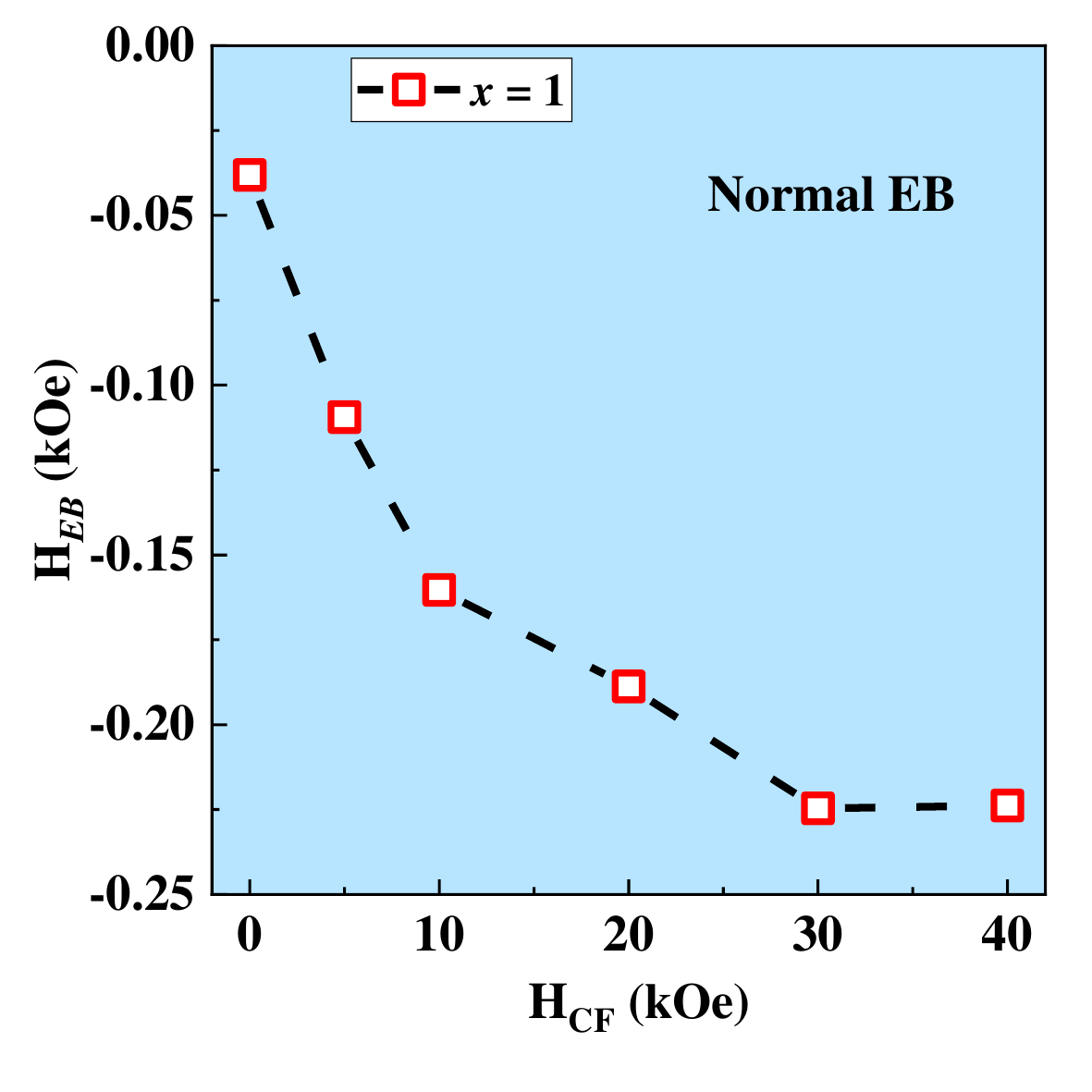}}
\caption{(Color online) Cooling-field dependence of the exchange-bias field ($H_{\mathrm{EB}}$) for $x = 1$.}
\label{fig:S13}
\end{figure}

\FloatBarrier

\begin{table}[H]
\caption{Phase fractions of the respective Mn valence states and average Mn valence obtained from XAS fitting.}
\label{tab:Mn_XAS}
\begin{ruledtabular}
\begin{tabular}{ccccc}
$x$ & Mn$^{2+}$ (\%) & Mn$^{3+}$ (\%) & Mn$^{4+}$ (\%) & Avg. Mn Valence \\ \colrule
0    & $12.88 \pm 5.30$ & $21.97 \pm 4.54$ & $65.15 \pm 3.03$ & $3.52 \pm 0.21$ \\
0.25 & $10.16 \pm 6.25$ & $15.62 \pm 5.47$ & $74.22 \pm 3.51$ & $3.64 \pm 0.25$ \\
0.50 & $15.83 \pm 6.11$ & $15.11 \pm 5.40$ & $69.06 \pm 3.60$ & $3.53 \pm 0.25$ \\
0.75 & $8.46 \pm 5.38$  & $16.15 \pm 5.00$ & $75.38 \pm 3.08$ & $3.67 \pm 0.22$ \\
1.00 & $14.39 \pm 7.55$ & $13.67 \pm 5.75$ & $71.94 \pm 2.88$ & $3.58 \pm 0.26$ \\
\end{tabular}
\end{ruledtabular}
\end{table}


\begin{table}[H]
\caption{Phase fractions of the respective Co valence states and average Co valence obtained from XAS fitting.}
\label{tab:Co_XAS}
\begin{ruledtabular}
\begin{tabular}{ccccc}
$x$ & Co$^{2+}$ HS (\%) & Co$^{3+}$ HS (\%) & Co$^{3+}$ LS (\%) & Avg. Co Valence \\ \colrule
0    & $90.91 \pm 0.45$ & $9.09 \pm 6.82$  & ---              & $2.09 \pm 0.20$ \\
0.25 & $72.13 \pm 3.67$ & $13.93 \pm 5.74$ & $13.93 \pm 7.34$ & $2.28 \pm 0.29$ \\
0.50 & $63.93 \pm 3.28$ & $20.49 \pm 5.34$ & $15.57 \pm 6.56$ & $2.36 \pm 0.26$ \\
0.75 & $56.91 \pm 3.66$ & $24.39 \pm 6.10$ & $18.70 \pm 6.91$ & $2.43 \pm 0.29$ \\
1.00 & $53.66 \pm 4.88$ & $29.27 \pm 8.13$ & $17.07 \pm 9.76$ & $2.46 \pm 0.39$ \\
\end{tabular}
\end{ruledtabular}
\end{table}


\begin{table}[H]
\caption{Oxygen deficiency ($\delta$) determined from XAS fitting and iodometric titration.}
\label{tab:delta}
\begin{ruledtabular}
\begin{tabular}{ccc}
$x$ & From XAS ($\delta$) & From titration ($\delta$) \\ \colrule
0    & $0.31 \pm 0.14$ & $0.50 \pm 0.01$ \\
0.25 & $0.15 \pm 0.18$ & --- \\
0.50 & $0.41 \pm 0.16$ & $0.45 \pm 0.01$ \\
0.75 & $0.54 \pm 0.16$ & --- \\
1.00 & $0.56 \pm 0.22$ & $0.69 \pm 0.01$ \\
\end{tabular}
\end{ruledtabular}
\end{table}


\begin{table}[H]
\caption{Average B-site valence determined from XAS fitting and iodometric titration.}
\label{tab:B_valence}
\begin{ruledtabular}
\begin{tabular}{ccc}
$x$ & From XAS & From titration \\ \colrule
0    & $2.81 \pm 0.15$ & $2.66 \pm 0.01$ \\
0.25 & $2.96 \pm 0.19$ & --- \\
0.50 & $2.95 \pm 0.18$ & $2.96 \pm 0.01$ \\
0.75 & $3.05 \pm 0.18$ & --- \\
1.00 & $3.02 \pm 0.24$ & $2.91 \pm 0.01$ \\
\end{tabular}
\end{ruledtabular}
\end{table}


\clearpage

\appendix

\section{EDAX Analysis}
\label{app:edax}
Table~\ref{tab:edax} summarizes the elemental compositions obtained from EDAX analysis. The compositions were normalized by fixing the Mn content to unity.

\begin{table}[H]
\caption{Nominal and EDAX-derived compositions of the NSCMO series. The values in parentheses represent the uncertainty in the last digits.}
\label{tab:edax}
\begin{ruledtabular}
\begin{tabular}{ccc}
$x$ & Nominal formula & Derived formula from EDAX \\
\hline
0 &
Nd$_2$CoMnO$_6$ &
Nd$_{2.01(14)}$Co$_{0.87(06)}$Mn \\

0.25 &
Nd$_{1.75}$Sr$_{0.25}$CoMnO$_6$ &
Nd$_{1.78(12)}$Sr$_{0.32(02)}$Co$_{0.91(06)}$Mn \\

0.50 &
Nd$_{1.50}$Sr$_{0.50}$CoMnO$_6$ &
Nd$_{1.55(11)}$Sr$_{0.53(04)}$Co$_{0.82(06)}$Mn \\

0.75 &
Nd$_{1.25}$Sr$_{0.75}$CoMnO$_6$ &
Nd$_{1.29(09)}$Sr$_{0.67(05)}$Co$_{0.84(06)}$Mn \\

1.00 &
NdSrCoMnO$_6$ &
Nd$_{1.06(07)}$Sr$_{0.93(07)}$Co$_{0.92(06)}$Mn \\
\end{tabular}
\end{ruledtabular}
\end{table}


\clearpage

\section{Iodometric Titration Methodology}
\label{app:iodometric}
\setlength{\parindent}{0pt}
\subsection{Chemistry of the Iodometric Titration Method}

The oxide powder was dissolved in an excess KI solution acidified with HCl. High-valent ions such as Co$^{3+}$/Co$^{4+}$ and Mn$^{3+}$/Mn$^{4+}$ oxidize I$^{-}$ to I$_2$ and are themselves reduced, typically to the +2 oxidation state:

\begin{equation}
2\,\mathrm{I}^{-}
\rightarrow
\mathrm{I}_{2}
+
2e^{-}.
\end{equation}

Each I$_2$ molecule generated in solution therefore corresponds to the transfer of two electrons to the transition-metal ions. The liberated iodine is subsequently titrated using a standard sodium thiosulfate solution with starch as an indicator:

\begin{equation}
\mathrm{I}_{2}
+
2\,\mathrm{S}_{2}\mathrm{O}_{3}^{2-}
\rightarrow
2\,\mathrm{I}^{-}
+
\mathrm{S}_{4}\mathrm{O}_{6}^{2-}.
\end{equation}

Thus, one I$_2$ molecule consumes two thiosulfate ions and accepts exactly two electrons. Since these reactions are stoichiometrically coupled, the total number of thiosulfate equivalents consumed is equal to the total number of electrons absorbed by Co and Mn ions in the sample. Sodium thiosulfate solutions are conventionally standardized in terms of normality ($\mathrm{eq\,L^{-1}}$), where one equivalent corresponds to one mole of electrons. Therefore, for a consumed volume $dV$ (L) of a solution with normality $C$, the total number of electron equivalents is

\begin{equation}
n=C\,dV.
\end{equation}

\subsection{Derivation of Oxygen Deficiency}

The oxygen deficiency ($\delta$) of the compound

\begin{equation}
\mathrm{Nd}_{2-x-s}\mathrm{Sr}_{x-p}\mathrm{Co}_{1-c}\mathrm{MnO}_{6-\delta}
\end{equation}

was determined by incorporating the experimentally measured deviations from nominal stoichiometry obtained from EDAX analysis, where $s$, $p$, and $c$ represent Nd-, Sr-, and Co-site deficiencies, respectively. The experimental quantities are defined as follows:

\begin{itemize}
\item $m$: sample mass (g)
\item $M$: molar mass of the compound (g mol$^{-1}$)
\item $C$: normality of Na$_2$S$_2$O$_3$ solution (eq L$^{-1}$)
\item $dV$: volume of Na$_2$S$_2$O$_3$ consumed (L)
\item $n$: total electron equivalents ($=C\,dV$)
\item $V_{\mathrm{Co}}$: average Co valence
\item $V_{\mathrm{Mn}}$: average Mn valence
\end{itemize}

Applying the charge-neutrality condition,

\begin{equation}
(2-x-s)\times3
+
(x-p)\times2
+
(1-c)V_{\mathrm{Co}}
+
V_{\mathrm{Mn}}
=
2(6-\delta).
\end{equation}

Expanding and simplifying,

\begin{align}
6-3x-3s+2x-2p+(1-c)V_{\mathrm{Co}}+V_{\mathrm{Mn}}
&=12-2\delta .
\end{align}

Therefore,

\begin{equation}
(1-c)V_{\mathrm{Co}}
+
V_{\mathrm{Mn}}
=
6+x+3s+2p-2\delta ,
\label{eq:charge_neutrality}
\end{equation}

Assuming complete reduction of Co and Mn to the +2 oxidation state, the number of electrons transferred per formula unit is

\begin{equation}
n_{\mathrm{mol}}
=
(1-c)(V_{\mathrm{Co}}-2)
+
(V_{\mathrm{Mn}}-2).
\end{equation}

Rearranging,

\begin{equation}
n_{\mathrm{mol}}
=
(1-c)V_{\mathrm{Co}}
+
V_{\mathrm{Mn}}
+
2c
-
4.
\end{equation}

Substituting Eq.~(\ref{eq:charge_neutrality}),

\begin{equation}
n_{\mathrm{mol}}
=
2+x+3s+2p+2c-2\delta .
\end{equation}

For a sample mass $m$,

\begin{equation}
n
=
\frac{m}{M}
\left(
2+x+3s+2p+2c-2\delta
\right).
\end{equation}

Rearranging for $\delta$,

\begin{equation}
2\delta
=
2+x+3s+2p+2c
-
\frac{Mn}{m},
\end{equation}

or

\begin{equation}
\delta
=
1
+
\frac{x}{2}
+
\frac{3s}{2}
+
p
+
c
-
\frac{Mn}{2m}.
\end{equation}

Since

\begin{equation}
n=C\,dV,
\end{equation}

the final expression for oxygen deficiency becomes

\begin{equation}
\boxed{
\delta
=
1
+
\frac{x}{2}
+
\frac{3s}{2}
+
p
+
c
-
\frac{M\,C\,dV}{2m}
}
\end{equation}

\end{document}